\documentclass[preprint,12pt,3p]{elsarticle}
\usepackage{amssymb}

\usepackage{graphicx}
\usepackage[utf8]{inputenc} 
\usepackage[T1]{fontenc}    
\usepackage{hyperref}       
\usepackage{url}            
\usepackage{booktabs}       
\usepackage{nicefrac}       
\usepackage{microtype}      
\usepackage{makecell}
\usepackage{amsfonts}       
\usepackage{color}
\usepackage{textcomp}
\usepackage{siunitx}
\usepackage{amsmath}
\usepackage{algorithm}
\usepackage{algorithmicx}
\usepackage{algpseudocode}
\usepackage{amssymb}
\usepackage{indentfirst}
\usepackage{subfigure}
\usepackage{multirow}
\usepackage[section]{placeins}
\usepackage{float}
\usepackage{xfrac}
\usepackage{ulem}
\usepackage{cancel}
\usepackage{hyperref}
\usepackage{changepage}
\usepackage{tabularx}
\usepackage{setspace}
\definecolor{dgreen}{RGB}{255,120,0}
\definecolor{myorange}{RGB}{255,153,0}
\journal{International Journal of Hydrogen Energy}

\begin{document}

\begin{frontmatter}

\title{Optimizing the Homogeneity and Efficiency of an SOEC Based on Multiphysics Simulation and Data-driven Surrogate Model}
\author[label1,label7,label8]{Yingtian Chi}
\author[label7]{Kentaro Yokoo}
\author[label7,label8]{Hironori Nakajima}
\author[label7,label8]{Kohei Ito}
\author[label1,label3]{Jin Lin\corref{cor2}}
\ead{linjin@tsinghua.edu.cn}
\author[label1,label2]{Yonghua Song}

\address[label1]{State Key Laboratory of Control and Simulation of Power Systems and Generation Equipment, Department of Electrical Engineering, Tsinghua University, Beijing 100087, China}
\address[label7]{Department of Hydrogen Energy Systems, Graduate School of Engineering, Kyushu University, Fukuoka, Japan}
\address[label8]{Department of Mechanical Engineering, Faculty of Engineering, Kyushu University, Fukuoka, Japan}
\address[label3]{Tsinghua-Sichuan Energy Internet Research Institute, Chengdu 610213, China}
\address[label2]{State Key Laboratory of Internet of Things for Smart City, University of Macau, Macau SAR 999078, China}

\cortext[cor2]{Corresponding author}

\address{}

\begin{abstract}
  Inhomogeneous current and temperature distributions are harmful to the durability of the solid oxide electrolysis cell (SOEC). 
  Segmented SOEC experiments reveal that a high steam utilization, which is favorable for system efficiency, leads to local steam starvation and enhanced the inhomogeneity. 
  It is necessary to consider inhomogeneity and efficiency jointly in optimization studies. 
  Three-dimensional (3D) multiphysics models validated with experiments can simulate the inhomogeneity in a reliable manner, but they are unsuitable for optimization due to the high computational cost. 
  This study proposes a method that combines segmented SOEC experiments, multiphysics simulation, and artificial intelligence to optimize the inhomogeneity and efficiency of SOEC jointly. 
  A 3D cell model is first built and verified by segmented SOEC experiments. 
  Then, fast neural network surrogate models are built from the simulation data and integrated into a multi-objective optimization problem. 
  Its solutions form a Pareto front reflecting the conflicting relationships among different objectives. 
  It is found that the down-stream current is 60\%-65\% of the up-stream current when the steam utilization is 0.7. 
  To increase the steam utilization to 0.8, the down-stream current will further drop to 50\%-60\% of the up-stream current. 
  The Pareto fronts enable system operators to achieve a balance between efficiency and inhomogeneity. 
\end{abstract}


\end{frontmatter}

\section{Introduction}
\label{sec:1}
{Solid oxide electrolysis cells (SOECs) constitute a promising energy storage technology for carbon neutralization owing to its high efficiency and ability to convert carbon dioxide into chemicals and fuels \cite{WANG2020-1},\cite{WANG2020},\cite{FRANK2018},\cite{KHANAFER2022},\cite{XU2021-2}.
However, the high operating temperature brings durability problems.
Mechanical stresses, induced by thermal transients and inhomogeneous temperature distributions, are the direct causes of various failures \cite{KHANAFER2022},\cite{Damm2006},\cite{Xu2021_1}.
The inhomogeneous distribution of current also contributes the cell degradation and reduces efficiency \cite{Schiller_2009}.
It is critical to improve the homogeneity of current and temperature for safety and durability.}

{Experimental studies that investigate the spatial inhomogeneity in SOFC and SOEC have been reported.
For measuring the temperature distribution, several methods have been reported.
Inserting thermocouples into the gas channels or the interconnectors is a widely-used method \cite{CANAVAR2016},\cite{RAZBANI2013},\cite{WU2022}.
Its spatial resolution is limited by the number of thermocouples, due to the difficulties in handling electrical wiring, insulation, and gas-tight sealing.
Guk et al. \cite{GUK2019} designed a multi-point thermal sensor to improve the spatial resolution of temperature distribution while keeping the wiring simple.
Other methods including infrared imaging \cite{SUGIHARA2021},\cite{SUGIHARA2020}, Raman spectroscopy \cite{Eigenbrodt2011}, and optical fiber sensors \cite{ZAGHLOUL2021}, have also been proposed to obtain high resolution temperature distribution data.}
{For measuring the current distribution, the segmented cell method is a powerful approach \cite{ZHANG2010},\cite{WANG2020-2}.
Wuillemin et al. \cite{wuillemin2008} tested an anode-supported planar SOFC with 18 cathode segmentations and measured the current-voltage (IV) curves and electrochemical impedance spectroscopy (EIS).
Using an anode-supported SOFC with 16 segmentations, Schiller et al. \cite{Schiller_2009} found that the inhomogeneity of current was enhanced remarkably under high fuel utilization due to the mass transport limitation occurring at the down-stream part.
They found that severe mass transport limitation induced inhomogeneous degradation.
Aydin et al. \cite{AYDIN2015} measured the current variation along an anode-supported microtubular SOFC with three segments.
It was observed that the current of the down-stream segment decreased rapidly due to fuel starvation as the voltage decreased, which could cause local redox cycles and lead to degradation.
They further found that the inhomogeneity was enhanced under methane operation due to the severe fuel starvation caused by incomplete methane reforming \cite{AYDIN2015-1}.
Kim et al. \cite{KIM2019} found that Ni re-oxidation and interface delamination tended to occur near the outlet in an anode-supported planar SOFC.
Wu et al. \cite{WU2022} tested an anode-supported SOFC with four cathode segments.
They found that high electrical loadings and high hydrogen dilution ratios could induce large temperature gradient and uneven current distribution, leading to harsh conditions near the gas outlet where microstructure degradation including cathode delamination and electrolyte crack were observed.
Therefore, the inhomogeneity of current density and temperature should be controlled for durability.}

To reduce experimental costs, multiphysics models are used to simulate the inhomogeneity, while their reliability can be verified by the experimental data \cite{WU2022},\cite{ZAGHLOUL2021},\cite{AYDIN2016}.
Bessler et al. used the data of a segmented anode-supported SOFC to calibrate a two-dimensional (2D) model \cite{Schiller_2009},\cite{Bessler2010}.
The result showed that the model accurately predicted the IV curves under different temperatures and gas compositions.
Aydin et al. used the data of a segmented tubular SOFC and a segmented planar SOFC to verify the corresponding numerical models \cite{AYDIN2016},\cite{AYDIN2018}.
However, the high computational cost of 3D multiphysics models hinders their applications, for example, in optimization studies.
To handle this problem, surrogate models have been constructed using simulation data produced by multiphysics models to predict the SOEC performances with low computational cost.
Arriagada et al. \cite{ARRIAGADA2002} and Huo et al. \cite{HUO2006} used artificial neural networks (ANNs) and support vector machine (SVM) to build SOFC surrogate models using simulation data of physical models.
Zahadat et al. \cite{ZAHADAT2015} and Milewski et al. \cite{MILEWSKI2009} used ANNs to predict the SOEC and SOFC performance under different operating parameters and design parameters from experimental data.
In our previous study, a method to build polynomial surrogate models is proposed \cite{CHI2020}.

{
For a fixed stack/system design, the operating parameters, including temperature, flow rates, gas compositions, voltage, and current, can be optimized to improve the efficiency and durability of SOEC systems.
Cai et al. \cite{CAI2014} studied the optimization of an SOEC system consisting of a stack and an air compressor with a one-dimensional (1D) SOEC model, seeking to maximize the efficiency and hydrogen production.
The temperature inhomogeneity was controlled by including a temperature gradient constraint.
Xing et al. \cite{XING2018} optimized the temperature, voltage, and steam flow rate of an SOEC system to maximize the hydrogen production under different target powers.
A 1D SOEC model was used and the temperature gradient constraint was also considered.
They found that the steam utilization affected the system efficiency significantly, because steam generation consumes a large amount of energy and it is hard to fully recover the heat contained in the unused steam.}
Their optimization results indicated that the optimal steam utilization to maximize system efficiency should exceed 80\%.
Such a high steam utilization will enhance the inhomogeneity of current and harm the durability as discussed above.
However, operation optimization studies that consider both temperature inhomogeneity and current inhomogeneity have not been reported, to the best of the authors' knowledge.
The reason is that operation optimization studies prefer to use lumped or 1D SOEC models due to the low computational cost.
It is difficult for such models to simulate the spatial inhomogeneity reliably due to the simplified geometry and boundary conditions.
3D models can simulate the cell performance with higher reliability, but they are too time-consuming for operation optimization studies \cite{XU2021-2}.
Additionally, most of the detailed spatial distributions simulated by 3D models are redundant for operation.

Combining experiments, multiphysics models, and fast surrogate models improve the reliability of multiphysics models and also enable multiphysics models to be integrated into optimization studies for fast numerical solutions \cite{XU2020},\cite{SUN2022}.
Xu et al. \cite{XU2020} built ANN surrogate models of SOFC using multiphysics simulation data, and used the surrogate models to optimize the power output under different fuel flow rates.
Using ANN surrogate models of SOFC, Sun et al. \cite{SUN2022} conducted optimization with multiple objectives including production rate, conversion rate, energy efficiency, and heat production.
In their study, the optimization problem was formulated to maximize the production rate while keeping the other objectives within given thresholds.
Xu et al. \cite{XU2021-2} optimized the operating parameters of an SOEC to maintain thermo-neutral operation.
From the optimization results, a four-dimensional map was constructed, which illustrated the relationships between voltage, temperature, power density, and gas composition under thermo-neutral operation.
With this map, it is convenient for the system operators to choose the operating parameters and ensure thermo-neutral operation.
However, the inhomogeneity of current and temperature was not considered in these studies, and the multiphysics models used by them were verified with only the overall IV curves. 

{This paper aims to combine segmented SOEC experiments, 3D multiphysics simulation, and neural network surrogate models to optimize the homogeneity and efficiency of SOEC. 
A 3D model of a cathode-supported SOEC is first built to predict the inhomogeneous distributions of current and temperature.
Its reliability is verified with the current and temperature distributions measured with a segmented SOEC.
With the simulation results, ANN surrogate models are trained to predict the current distribution and temperature distribution under different cell voltages, flow rates, and temperatures.
A multi-objective optimization problem is formulated with the surrogate models to optimize the homogeneity, temperature, steam utilization, voltage, and hydrogen production under different electrolysis powers.
The multi-objective optimization problem is solved by decomposing the original problem into a series of single objective optimization problems, whose solutions form the Pareto front reflecting the conflicting relationship among different objectives.}
The results show that the homogeneity, cell voltage, and hydrogen production accord with each other, while a high steam utilization degrades the homogeneity of current.
If the steam utilization is 0.7, the current of the down-stream segment will be 60\%-65\% of that of the up-stream segment.
To increase the steam utilization to 0.8, the current of the down-stream segment will drop to 50\%-60\% of that of the up-stream segment. 
The optimal solution is chosen from the Pareto front to achieve a trade-off between different objectives.
The Pareto fronts can be easily delivered to system operators by stack manufacturers in product datasheets, so that the former can choose the optimal operating point to balance between system efficiency and inhomogeneity.
In this way, the collaboration between stack manufacturers and system operators can be enhanced.

The main contributions of this study include:
(1) Segmented SOEC experiments, 3D multiphysics simulation, and artificial intelligence are combined to build fast surrogate models of an SOEC that can predict the electrical performance and spatial inhomogeneity of the SOEC.
With this method, the distribution of current density is considered in operation optimization quantitatively.
(2) Pareto fronts that reflect the conflicting relationships between inhomogeneity, temperature, hydrogen production, and steam utilization are constructed.
It enables system operators to choose the optimal operating parameters, including the temperature, the voltage, and the steam flow rate, to balance between inhomogeneity and efficiency.

This paper is organized as follows.
Section \ref{sec:2} introduces the segmented SOEC experiment and the multiphysics SOEC model.
Section \ref{sec:3} introduces the training of the neural network surrogate models, as well as how to formulate and solve the multi-objective optimization problems.
Section \ref{sec:4} discusses the validation of the models, as well as the solutions of the multi-objective optimization problems.
\FloatBarrier
\section{Segmented-electrode experiments and multiphysics model}
\FloatBarrier
\label{sec:2}
\subsection{Overview of the study}
\begin{figure}
  \centering
  \includegraphics[width=1\textwidth]{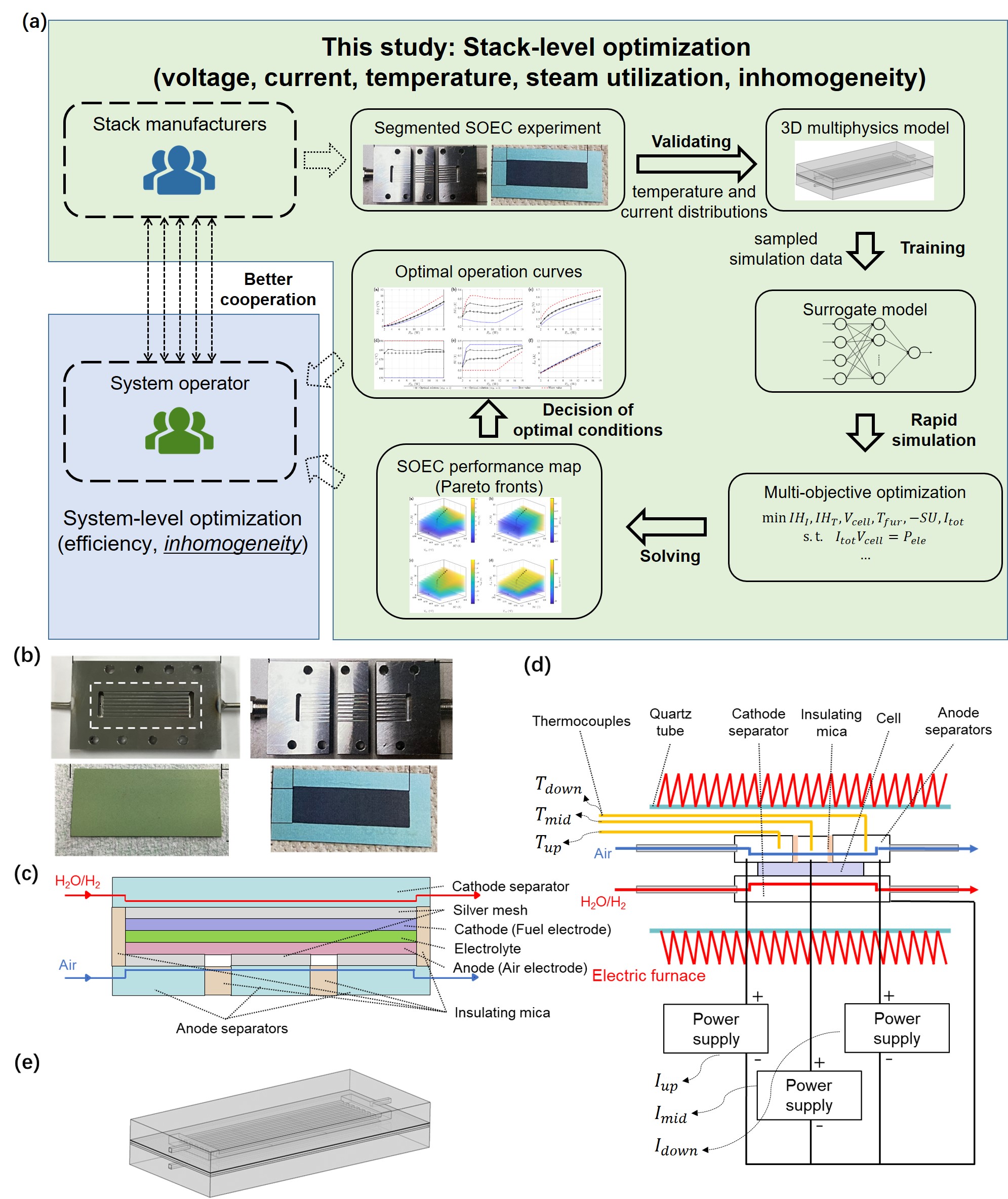}
  \caption{(a) Overview of this study. (b) The SOEC and segmented separators. (c) Cross-section of the cell assembly. (d) Schematic diagram of the test rig. (e) Geometry of the 3D multiphysics model.}
  \label{fig:studyoverview}
\end{figure}

\FloatBarrier
{An overview of this study is provided in Figure \ref{fig:studyoverview}(a).
Experiments are conducted on a cathode-supported SOEC to measure the distributions of temperature and current density under different cell voltages and steam flow rates.
To reduce the experimental cost, the limited experimental data are used to validate a 3D multiphysics model, which can simulate the SOEC performance with high reliability.
However, the high computational cost of the 3D model hinders its application in optimization studies. 
To solve this problem, fast artificial neural network (ANN) surrogate models are trained with the simulation data generated by the 3D model.
They can rapidly simulate the distributions of the cell characteristics, including temperature and current density.
Therefore, the surrogate models are used for a multi-objective optimization problem to optimize objectives including the cell voltage, temperature, hydrogen production, steam utilization, and inhomogeneity.}
The problem is solved numerically, and the Pareto fronts reflecting the SOEC performances are constructed, which reveals the conflicting relationships between multiple objectives.
Optimal solutions are chosen on the Pareto fronts, which form optimal operation curves which can achieve a trade-off among conflicting objectives. 
The optimization results can be parameterized and delivered to the system operators in product datasheets by the stack manufacturers, so that system operators can choose the operating conditions to balance system efficiency and inhomogeneity.
In the following part of this section, the segmented SOEC experiment and the 3D multiphysics model are introduced.

\subsection{Segmented SOEC experiment}

{Figure \ref{fig:studyoverview}(b) shows the segmented cell used in the experiment.
The cell was a commercial cathode-supported SOEC (ASC-10B, Elcogen AS), while the anode and cathode separators were designed and manufactured in-house.
The geometric parameters of the cell are listed in Table \ref{tab:cell}.
Figure \ref{fig:studyoverview}(c) demonstrates the cross section of the cell assembly.
Silver meshes were attached to the cathode and anode for current collection.
On the cathode side, only one piece of silver mesh was attached, covering the active area, i.e., the area opposite to the anode.
On the anode side, three pieces of silver meshes were attached, covering the up-stream segment, middle-stream segment and down-stream segment of the anode, respectively.
The meshes were carefully tailored to avoid short circuiting.
Silver paste was applied between the meshes and the electrodes for reliable electrical contact.
The cell was assembled with the metallic separators.
The gaps between the separator and cell were sealed by mica sheets and ceramic sealant (Alon Ceramic, Shin-Etsu Chemical).
The anode separators were also divided into three segments, corresponding to the up-stream segment, middle-stream segment and down-stream segment, respectively.
The gaps between the three anode separators were sealed and insulated by mica sheets and ceramic sealant (Alon Ceramic, Shin-Etsu Chemical).}

{Figure \ref{fig:studyoverview}(d) illustrates the schematic diagram of the test rig.
The cell assembly was placed in a quartz tube and inserted into a tubular furnace.
Heat insulation wool was stuffed into both ends of the quartz tube to reduce heat dissipation.
Air and a mixture of steam and hydrogen were fed to the anode side and the cathode side, respectively.
The hydrogen/steam mixture was generated by a bubbler with hydrogen being the carrier gas.
The temperature of the bubbler was controlled, and the steam partial pressure was calculated according to the saturation vapor pressure.
The pipe between the bubbler and the cell was wound with electrical heating band to avoid water condensation.
Three thermocouples were inserted into the anode separators to measure the temperature distribution.
The positive electrodes of three power supplies were connected to the three silver meshes on the SOEC anode, while the negative electrodes were connected to the silver mesh on the SOEC cathode.
The voltages of the three power supplies were kept the same to reproduce an actual planar SOEC without segmentation.
Since the resistance of silver mesh was significantly lower than the in-plane resistance of the anode, it was assumed that there is no in-plane current in the anode.
Therefore, the current distribution can be measured with the three power supplies.
IV curves and the temperature variations were measured under the experimental conditions listed in Table \ref{tab:expcondition}.}

\begin{table}
    \footnotesize
    \caption{Geometric parameters of the SOEC.}
     \centering
     \begin{tabular}{cccc}
       \toprule
        Name&Value&Name &Value \\
       \midrule
       Anode thickness& \SI{25}{\micro\meter} & Electrolyte thickness& \SI{3}{\micro\meter} \\
       Cathode thickness& \SI{500}{\micro\meter} & Cell length& \SI{65}{\milli\meter}\\
       Cell width& \SI{30}{\milli\meter}& Anode width& \SI{15}{\milli\meter}\\
       Anode length& \SI{48}{\milli\meter}& Channel height& \SI{1}{\milli\meter}\\
       Separator length& \SI{85}{\milli\meter}& Separator width& \SI{50}{\milli\meter}\\
       Separator thickness& \SI{11}{\milli\meter}& &\\
       \bottomrule
     \end{tabular}
     \label{tab:cell}
\end{table}
\begin{table}
    \footnotesize
    \caption{Experimental conditions. (sccm: \SI{25}{\celsius}, 1 atm)}
     \centering
     \begin{tabular}{ccccc}
       \toprule
        Condition& Temperature & Air flow rate & steam flow rate& hydrogen flow rate \\
       \midrule
       1& \SI{660}{\celsius}&\SI{400}{sccm}&\SI{40}{sccm}&\SI{40}{sccm}\\
       2& \SI{660}{\celsius}&\SI{400}{sccm}&\SI{120}{sccm}&\SI{120}{sccm}\\
       \bottomrule
     \end{tabular}
     \label{tab:expcondition}
\end{table}

\subsection{The 3D multiphysics SOEC model}
\label{sec:3Dmodel}

The geometry of the 3D multiphysics model is shown in Figure \ref{fig:studyoverview}(e).
Instead of a repeating channel unit, the whole cell assembly is modelled, which includes the SOEC and the metallic separators, because the heat transfer between the cell assembly and the furnace environment significantly influences the temperature distribution. 
Only the representative equations are introduced, as listed in Table \ref{tab:pde}, because they can be found in most 3D modeling studies of SOEC.

{The mass and momentum transfer processes are defined in the gas channels and the porous electrodes.
Equation (\ref{eqn:MassGov}) describes the mass conservation.
The Maxwell-Stefan diffusion model, consisting of Equations (\ref{eqn:MassO2Gov})-(\ref{eqn:MassH2Gov2}), describes the mass balance of species.
Equation (\ref{eqn:MomentumGov}), the Navier-Stokes equation, models the momentum conservation.
$\varepsilon$ and $\kappa$ are the porosity and permeability of the porous media, respectively.
It is assumed that $\varepsilon=1$ and $\kappa=+\infty$ in the channels.
The energy balance equation is defined in all of the domains, formulated as Equation (\ref{eqn:EnergyGov}).
The charge transfer equations are defined in the electrolyte and the electrodes, formulated as Equations (\ref{eqn:Charge3DGov})-(\ref{eqn:Nernst}).
Equation (\ref{eqn:Charge3DGov}) is the charge conservation law applied in different domains.
$\sigma$ is the conductivity, $\mathbf{i}$ is the current density vector, $\phi$ is the potential, and $S_{\rm act}$ is the volumetric density of active area.
$i_{\rm a}$ and $i_{\rm c}$ are the local electrochemical current sources.
The subscripts "ion", "elec", "a", "c", and "ele" represent the ionic phase, electronic phase, anode, cathode, and electrolyte, respectively.
Equation (\ref{eqn:CurrentContinue}) describes the continuity of current density vector and potential on the interfaces between neighboring domains.
The ionic phase potential and the electronic phase potential are related to each other through the activation overpotentials, $\eta_{\rm c}$ and $\eta_{\rm a}$, and the reversible potentials, $E_{\rm ref,c}$ and $E_{\rm ref,a}$, as shown in Equation (\ref{eqn:Charge2D3DRelation}).
}
The source terms in the governing equations are listed in Table \ref{tab:source}.
The material parameters are adopted from the 3D model of a cathode-supported SOEC (Elcogen AS) reported by Wehrle at al. \cite{WEHRLE2022}.
The exchange current densities are tuned with experiment data.

\begin{table}[H]
  \footnotesize
  \caption{{Governing equations for the 3D model.}}
   \centering
\begin{tabularx}{\textwidth}{p{3cm} X<{\centering}}
  \toprule
        &Governing equations\\
       \midrule
       \makecell[c]{Mass balance}& 
       \begin{equation}
        \label{eqn:MassGov}
        \nabla\cdot\left(\rho\mathbf{u}\right)=Q_{\rm mass}
      \end{equation}

      \begin{equation}
        \label{eqn:MassO2Gov}
        \nabla\cdot\left(\mathbf{j}_{\rm i}+\rho\omega_{\rm i}\mathbf{u}\right)=Q_{\rm mass,i},\quad {\rm i=H_2,O_2}
      \end{equation}

        \begin{equation}
          \label{eqn:MassO2Gov1}
          \mathbf{j}_{\rm i}=-\rho\omega_{\rm i}\left(\tilde{D}_{\rm i,i}\mathbf{d}_{\rm i}+\tilde{D}_{\rm i,j}\mathbf{d}_{\rm j}\right),\quad {\rm (i,j)=(H_2,H_2O),(O_2,N_2)}
        \end{equation}

          \begin{equation}
            \label{eqn:MassH2Gov2}
            \mathbf{d}_{\rm i}=\nabla x_{\rm i}+\frac{\nabla p}{p}\left(x_{\rm i}-\omega_{\rm i}\right),\quad{\rm i=H_2,O_2,H_2O,N_2}
          \end{equation}
          \\
        \makecell[c]{Momentum balance}&
          \begin{equation}
            \label{eqn:MomentumGov}
            \begin{aligned}
            \frac{1}{\varepsilon}\rho \mathbf{u}\cdot\nabla\mathbf{u}=-\varepsilon\nabla p&+\nabla\cdot\left(\mu\left(\nabla\mathbf{u}+(\nabla\mathbf{u})^{\rm T}\right)-\frac{2}{3}\mu\nabla\cdot\mathbf{u}\right)-\left(\frac{\varepsilon\mu}{\kappa}+\frac{Q_{\rm mass}}{\varepsilon}\right)\mathbf{u}
            \end{aligned}
          \end{equation}
      \\
      \makecell[c]{Energy balance}&
      \begin{equation}
        \label{eqn:EnergyGov}
        \nabla\cdot\left(-\lambda\nabla T\right)+\rho c_{\rm p}\mathbf{u}\nabla T=Q_{\rm heat}
      \end{equation}
      
      \\
      \makecell[c]{Charge balance}&
       \begin{equation}
        \label{eqn:Charge3DGov}
        \left\{
          \begin{aligned}
            &\nabla\cdot\left(-\sigma_{\rm ion,a}\nabla\phi_{\rm ion,a}\right)\equiv\nabla\cdot \mathbf{i}_{\rm ion,a}=S_{\rm act}i_{\rm a}&,{\rm anode}\\
            &\nabla\cdot\left(-\sigma_{\rm elec,a}\nabla\phi_{\rm elec,a}\right)\equiv\nabla\cdot \mathbf{i}_{\rm elec,a}=-S_{\rm act}i_{\rm a}&,{\rm anode}\\
            &\nabla\cdot\left
            (-\sigma_{\rm ion,c}\nabla\phi_{\rm ion,c}\right)\equiv\nabla\cdot \mathbf{i}_{\rm ion,c}=-S_{\rm act}i_{\rm c}&,{\rm cathode}\\
            &\nabla\cdot\left(-\sigma_{\rm elec,c}\nabla\phi_{\rm elec,c}\right)\equiv\nabla\cdot \mathbf{i}_{\rm elec,c}=S_{\rm act}i_{\rm c}&,{\rm cathode}\\
            &\nabla\cdot\left(-\sigma_{\rm ion,ele}\nabla\phi_{\rm ion,ele}\right)\equiv\nabla\cdot \mathbf{i}_{\rm ion,ele}=0&,{\rm electrolyte}
          \end{aligned}
        \right.
      \end{equation} 
      \begin{equation}
        \label{eqn:CurrentContinue}
        \left\{
          \begin{aligned}
            &\mathbf{i}_{\rm ion,a}=\mathbf{i}_{\rm ion,ele},\phi_{\rm ion,a}=\phi_{\rm ion,ele}&,{\rm anode|electrolyte\ interface}\\
            &\mathbf{i}_{\rm ion,c}=\mathbf{i}_{\rm ion,ele},\phi_{\rm ion,c}=\phi_{\rm ion,ele}&,{\rm cathode|electrolyte\ interface}
          \end{aligned}
        \right.
      \end{equation}
      \begin{equation}
        \label{eqn:Charge2D3DRelation}
        \left\{
          \begin{aligned}
        \eta_{\rm c}&=\phi_{\rm elec,c}-\phi_{\rm ion,c}-E_{\rm ref,c}\\
        \eta_{\rm a}&=\phi_{\rm elec,a}-\phi_{\rm ion,a}-E_{\rm ref,a}
      \end{aligned}
      \right.
      \end{equation}
\begin{equation}
  \label{eqn:Nernst}
  \left\{
    \begin{aligned}
  E_{\rm ref,c}&=\frac{RT}{2F}\log\frac{x_{\rm H_2O}}{x_{\rm H_2}}\\
  E_{\rm ref,a}&=E_{\rm ref,0}+\frac{RT}{4F}\log\frac{x_{\rm O_2}p_{\rm a}}{p_{\rm atm}}
    \end{aligned}
  \right.
\end{equation}
      \\
       \bottomrule
\end{tabularx}
\label{tab:pde}
\end{table}

\begin{table}
  \footnotesize
  \caption{Source terms in the governing equations.}
   \centering
\begin{tabularx}{\textwidth}{p{5cm} X<{\centering}}
  \toprule
        &Source terms\\
       \midrule
       \makecell[c]{Mass source}&
       \begin{equation}
        \label{eqn:BC_masssource}
        Q_{\rm mass}=\left\{
          \begin{aligned}
        &\frac{i_{\rm a}M_{\rm O_2}}{4F}&,{\rm anode}\\
        &\frac{i_{\rm c}(M_{\rm H_2}-M_{\rm H_2O})}{2F}&,{\rm cathode}\\
        &0&,{\rm channels}
          \end{aligned}
        \right.
      \end{equation}
      \begin{equation}
        \label{eqn:BC_masssourcei}
        Q_{\rm mass,O_2}=\left\{
          \begin{aligned}
        &\frac{i_{\rm a}M_{\rm O_2}}{4F}&,{\rm anode}\\
        &0&,{\rm anode\ channels}
          \end{aligned}
        \right.
      \end{equation}
      \begin{equation}
        \label{eqn:BC_masssourcei1}
        Q_{\rm mass,H_2}=\left\{
          \begin{aligned}
        &\frac{i_{\rm c}M_{\rm H_2}}{2F}&,{\rm cathode}\\
        &0&,{\rm cathode\ channels}
          \end{aligned}
        \right.
      \end{equation} 
      \\
      \makecell[c]{Heat source}&
\begin{equation}
  \label{eqn:BC_elecheatsource}
  Q_{\rm heat}=\left\{
    \begin{aligned}
      &\frac{||\mathbf{i}_{\rm ion,ele}||^2}{\sigma_{\rm ion,ele}}&,{\rm electrolyte}\\
      &i_{\rm c}\eta_{\rm c}+\frac{||\mathbf{i}_{\rm ion,c}||^2}{\sigma_{\rm ion,c}}+\frac{||\mathbf{i}_{\rm elec,c}||^2}{\sigma_{\rm elec,c}}&,{\rm cathode}\\
      &i_{\rm a}\eta_{\rm a}+\frac{||\mathbf{i}_{\rm ion,a}||^2}{\sigma_{\rm ion,a}}+\frac{||\mathbf{i}_{\rm elec,a}||^2}{\sigma_{\rm elec,a}}+\frac{i_{\rm a}T\Delta S_{\rm r}}{2F}&,{\rm anode}\\
      &0&,{\rm others}
    \end{aligned}
  \right.
\end{equation}
      \\
      \makecell[c]{Electrochemical current source}&
\begin{equation}
  \label{eqn:BV}
  \left\{
    \begin{aligned}
  i_{\rm c}&=i_{\rm 0,c}\left(\frac{x_{\rm H_2}}{x_{\rm H_2,ref}}\exp\left(\frac{\alpha F\eta_c}{RT}\right)-\frac{x_{\rm H_2O}}{x_{\rm H_2O,ref}}\exp\left(-\frac{\alpha F\eta_c}{RT}\right)\right)\\
  i_{\rm a}&=i_{\rm 0,a}\left(\frac{x_{\rm O_2}}{x_{\rm O_2,ref}}\exp\left(\frac{\alpha F\eta_a}{RT}\right)-\exp\left(-\frac{\alpha F\eta_a}{RT}\right)\right)
\end{aligned}
\right.
\end{equation}
      \\
       \bottomrule
\end{tabularx}
\label{tab:source}
\end{table}

\section{Multi-objective optimization based on surrogate models}
\label{sec:3}
\subsection{Construction of artificial neural network surrogate models}
\label{sec:3-1}
\begin{figure}
  \centering
  \includegraphics[width=1\textwidth]{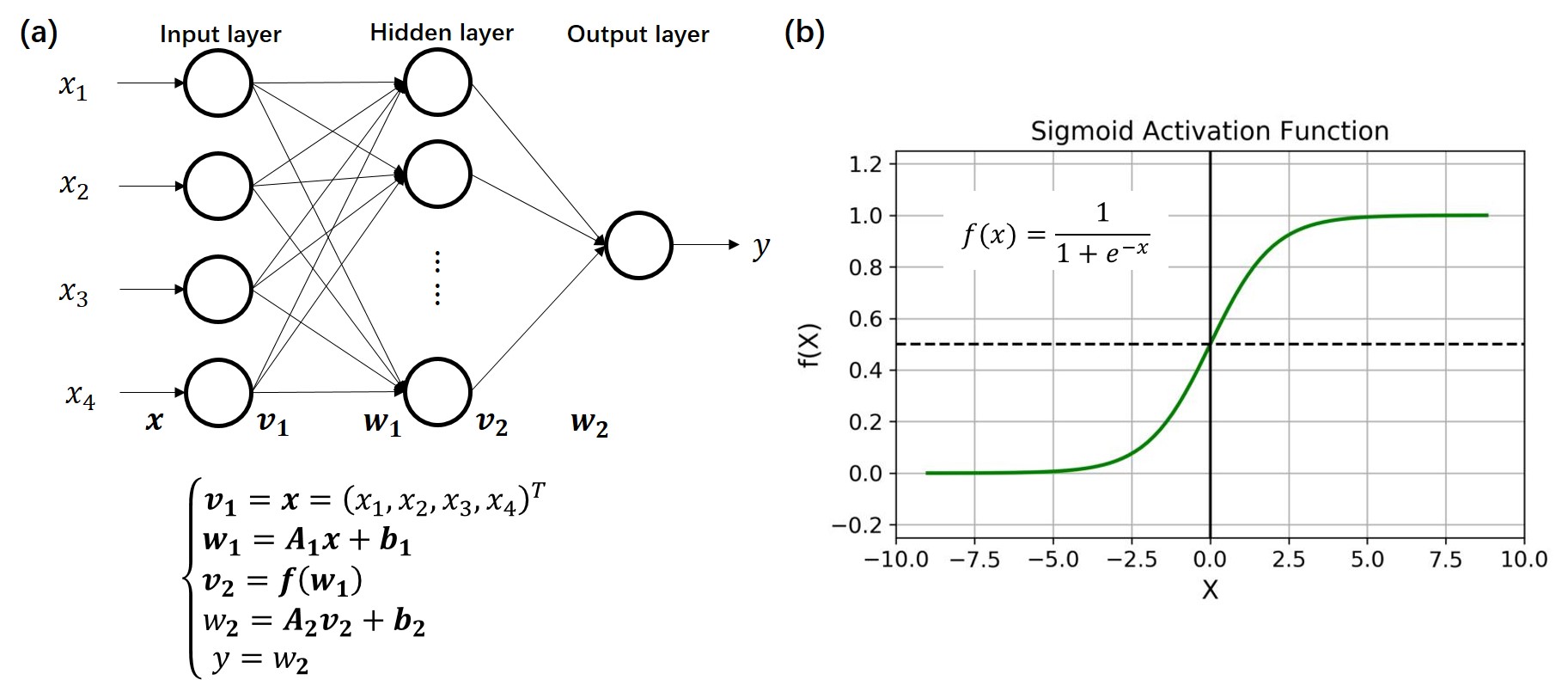}
  \caption{(a) Graphical illustration of the artificial neural network. (b) Illustration of the sigmoid activation function.}
  \label{fig:NNstruct}
\end{figure}

{The 3D multiphysics model is unsuitable for optimization due to its high computational cost.
This study handles this problem by building artificial neural network (ANN) surrogate models from the simulation data of the 3D model.
The ANN models can predict the SOEC performances with low computational cost, which enables the numerical solution of the optimization problem.

Figure \ref{fig:NNstruct}(a) is a graphical illustration of the ANN model.
An ANN model consists of neurons connected by branches.
The neurons and branches represent nonlinear operations and linear operations, respectively.
In this study, the neurons adopt the sigmoid function shown in Figure \ref{fig:NNstruct}(b), which is called the activation function.
It enables the ANN model to approximate complex nonlinear functions.
The numbers of neurons and layers in the ANN model are determined using the trial-and-error method.
} 

{Before building the ANN models, the inputs and the outputs are determined first.
Four inputs and five outputs are chosen, as shown in Table \ref{tab:simulationdatabase}.
These outputs are used to calculate the performance indices of the SOEC, as defined in Equation (\ref{eqn:unindex}).
$IH_{I}$ and $IH_{T}$ quantify the inhomogeneity of the distributions of current and temperature.
A low $IH_{I}$ and a low $IH_{T}$ are beneficial for the durability of the SOEC.
$SU$ is the steam utilization.
A high $SU$ is necessary to improve the system efficiency.}
\begin{equation}
  \label{eqn:unindex}
  \left\{
      \begin{aligned}
          IH_{I}&=1-\frac{I_{\rm down}}{I_{\rm up}}\\
          IH_{T}&=T_{\rm max}-T_{\rm min}\\
          SU&=\frac{I_{\rm up}+I_{\rm mid}+I_{\rm down}}{2FQ_{\rm st}}
      \end{aligned}
  \right.
  \end{equation}

Five sets of the ANN surrogate model in Figure \ref{fig:NNstruct}(a), with 10, 10, 10, 5, and 5 neurons in the hidden layer, are built to predict the five outputs listed in Table \ref{tab:simulationdatabase}, respectively.
For accurate prediction, the weights in the ANN models, represented by the branches in Figure \ref{fig:NNstruct}(a), should be adjusted, which is known as training.
The 3D model is simulated under 1764 sets of inputs sampled randomly within the ranges listed in Table \ref{tab:simulationdatabase} to generate 1764 data points.
1500 of them are used to train the ANN models, called the training dataset.
The training process is realized by minimizing the fitting errors on the training dataset using the Levenberg-Marquardt algorithm.
The rest of the 1764 data points are used to test the accuracy of the ANN models, called the test dataset.

\begin{table}[H]
  \footnotesize
  \caption{Inputs and outputs of the ANN surrogate models.}
   \centering
   \begin{tabular}{cccc}
     \toprule
      \multicolumn{2}{c}{Variables} & Meaning&Range\\
      \midrule
      \multirow{4}{*}{Input} &$T_{\rm fur}$&Furnace temperature & [600,750]\SI{}{\celsius}\\
      &$Q_{\rm air}$&Air flow rate&[40,300]\SI{}{sccm}\\
      &$Q_{\rm st}$&Steam flow rate&[20,150]\SI{}{sccm}\\
      &$V_{\rm cell}$&Cell voltage&[1.0,1.7]\SI{}{V}\\
      \midrule
      \multirow{5}{*}{Output} &$T_{\rm max}$&Maximum temperature & \\
      &$T_{\rm min}$&Minimum temperature & \\
      &$I_{\rm up}$&Current of the up-stream segment&\\
      &$I_{\rm mid}$&Current of the middle-stream segment&\\
      &$I_{\rm down}$&Current of the down-stream segment&\\
     \bottomrule
   \end{tabular}
   \label{tab:simulationdatabase}
\end{table}

\subsection{Sensitivity analysis}
\label{sec:3-2}
{Before solving the optimization problem, the global sensitivity of $IH_{I}$, $IH_{T}$, and $SU$ against the inputs ($Q_{\rm st}$, $Q_{\rm air}$, $T_{\rm fur}$, and $V_{\rm cell}$) is evaluated with the Sobol sensitivity analysis, in order to determine which of the inputs are worth being adjusted to optimize the cell's performance indices.

The Sobol sensitivity is a variance-based sensitivity index, which quantifies to which extent the inputs affect the output \cite{YAN2019},\cite{SALTELLI2010}.
Assume that $Y$ is an output variable affected by $k$ input variables which are denoted as $X_1$,...,$X_k$.
The Sobol analysis assumes that $Y$ can be decomposed as Equation (\ref{eqn:SobolDecomp}).
Accordingly, the variance of $Y$ can be decomposed into the sum of the partial variances of individual inputs and the partial variances of the interaction terms, as shown in Equation (\ref{eqn:SobolDecompVar}).
With this decomposition, a first-order sensitivity index $S_i$ and a total-effect sensitivity index $ST_i$ are defined as Equation (\ref{eqn:SobolSens}).
The former quantifies the first-order contribution of $X_i$ to the output variance, while the latter quantifies all the contributions related with $X_i$ (including the interaction terms that contain $X_i$) to the output variance.
The larger $S_i$ and $ST_i$ are, the more the output is influenced by the $i$th input.
In this study, the Sobol sensitivity is estimated by quasi Monte-Carlo simulation with the ANN surrogate models using the algorithm reported by Saltelli et al. \cite{SALTELLI2010}, and the results are presented in Section \ref{sec:4-2}.}
\begin{equation}
  \label{eqn:SobolDecomp}
  Y=f_0+\sum_{i=1}^k f_i(X_i)+\sum_{i=1}^k \sum_{j=i+1}^k f_{i,j}(X_i,X_j)+\cdots+f_{1,2,\cdots,k}(X_1,X_2,\cdots,X_k)
\end{equation}
\begin{equation}
  \label{eqn:SobolDecompVar}
  Var(Y)=\sum_{i=1}^k V_i+\sum_{i=1}^k \sum_{j=i+1}^k V_{i,j}+\cdots+V_{1,2,\cdots,k}
\end{equation}
\begin{equation}
  \label{eqn:SobolSens}
  \left\{
    \begin{aligned}
      S_i&=\frac{V_i}{Var(Y)}\\
      ST_i&=\frac{V_i+\sum_{j=i+1}^k V_{i,j}+\cdots+V_{1,2,\cdots,k}}{Var(Y)}
    \end{aligned}    
  \right.
\end{equation}

\subsection{Multi-objective optimization to improve the efficiency and inhomogeneity}
\label{sec:3-3}
{It is a common situation that there are multiple objectives conflicting mutually in an actual industrial system \cite{SHARIFZADEH2017}.
For example, Yan et al. \cite{YAN2019} found that there was a trade-off between the electrochemical performance and lifetime of SOFC cathode when optimizing microstructure parameters.
Minimization of one objective would inevitably increase the other objective.
}

{
When integrated with fluctuating renewable energies, the electrolysis power of the SOEC is also varying.
Thus, it is favorable that the efficiency is maximized and the inhomogeneity is minimized under different input powers \cite{WU2022},\cite{AYDIN2016}.
The inhomogeneity is quantified by the inhomogeneity indices of temperature and current, $IH_{I}$ and $IH_{T}$, defined in Equation (\ref{eqn:unindex}).
Since the SOEC system parameter is unavailable in this study, the electrolysis current $I_{\rm tot}$ (proportional to the hydrogen production $Q_{\rm H_2}$), the cell voltage $V_{\rm cell}$, the temperature $T_{\rm fur}$, the steam utilization $SU$ are chosen as the objectives.
These stack-level variables are closely related to the system efficiency, and they can also be determined by stack manufacturers without the knowledge of system structures.
Apparently, these objectives cannot fulfill their optimal values simultaneously.
For example, increasing $SU$ is beneficial for efficiency but leads to a large concentration gradient along the flow direction.
Consequently, the inhomogeneity of current is enhanced, meaning that $IH_{I}$ also increases.}
{The conflicting multiple objectives can be handled by multi-objective optimization.
In this study, a multi-objective optimization problem considering six objectives are formulated, as defined in Equation (\ref{eqn:multi_op}).
This optimization problem aims to minimize the inhomogeneity ($IH_{I}$ and $IH_{T}$) and maximize the efficiency with the electrolysis power equals to a specified value, $P_{\rm ele}$.
There are negative signs on $SU$ and $I_{\rm tot}$ because they are positively correlated with the efficiency while this problem is a minimization problem. 
In addition to the constraint of $P_{\rm ele}$, the constraints also require that the decision variables are kept within the ranges listed in Table \ref{tab:simulationdatabase}.
This problem will be solved under different $P_{\rm ele}$ to find the optimal operating conditions under different power consumptions.
}

\begin{subequations}
    \label{eqn:multi_op}
    \begin{equation}
        \min_{Q_{\rm st},Q_{\rm air},T_{\rm fur},V_{\rm cell}} IH_{I},IH_{T},V_{\rm cell},T_{\rm fur},-SU,-I_{\rm tot}
    \end{equation}
    \begin{equation}
      {\rm s.t.}\quad V_{\rm cell}I_{\rm tot}=P_{\rm ele}
    \end{equation}
    \begin{equation}
        V_{\rm cell,min}\le V_{\rm cell}\le V_{\rm cell,max}
    \end{equation}
    \begin{equation}
        Q_{\rm st,min}\le Q_{\rm st}\le Q_{\rm st,max}
    \end{equation}
    \begin{equation}
        Q_{\rm air,min}\le Q_{\rm air}\le Q_{\rm air,max}
    \end{equation}
    \begin{equation}
        T_{\rm fur,min}\le T_{\rm fur}\le T_{\rm fur,max}
    \end{equation}
\end{subequations}

{
In order to solve the multi-objective problem, the Pareto optimality condition \cite{SHARIFZADEH2017} is adopted in this study.
It states that, for any Pareto-optimal solution, it is impossible to find another solution that dominates this solution on all of the objectives.
In other words, for any Pareto-optimal solution, it is impossible to improve one of the objective without sacrificing any other objectives.
This definition allows the existence of multiple optimal solutions, which form a solution set called the Pareto front.
The Pareto front can reflect the relationships between multiple conflicting objectives.}

The optimization problem ise nonlinear owing to the ANN surrogate models.
{There are several methods to solve such multi-objective problems.
Heuristic algorithms were reported to solve multi-objective problems and construct the Pareto fronts.
Yan et al. \cite{YAN2019} used non-dominated sorting genetic algorithm (NSGA-II) to optimize the synthesis conditions of SOFC cathodes to improve lifetime and electrochemical performance.
Behzadi et al. \cite{BEHZADI2021} and Lei et al. \cite{LEI2022} used the grey-wolf algorithm and the multi-objective particle swarm optimization algorithm to optimize the cost and the efficiency of SOFC systems, respectively.
Sharifzadeh et al. \cite{SHARIFZADEH2017} optimized the profitability and the safe operating range of an SOFC power generation system.
By applying different weights to the two objectives, they transformed the multi-objective problem into a series of single-objective problems, and the solutions formed the Pareto front.
}

In this paper, the brute-force grid search method is adopted, which is simple and effective \cite{YAN2019}.
First, $T_{\rm fur}$ and $SU$ are discretized into several levels within $[600,750]^\circ{\rm C}$ and $[0.5,0.9]$, respectively, and a grid is formed.
At each grid point $(T_{\rm fur,i},SU_{\rm j})$, the original multi-objective optimization problem is transformed into Equation (\ref{eqn:multi_op_equi}).
The constraints in Equation (\ref{eqn:multi_op}) still apply.
It will be discussed in Section \ref{sec:4-4} that the solution of Equation (\ref{eqn:multi_op_equi}) will form the solution of the original problem, Equation (\ref{eqn:multi_op}).
Additionally, it will be proven in Section \ref{sec:4-4} that the four objectives in Equation (\ref{eqn:multi_op_equi}) accord with each other.
Therefore, Equation (\ref{eqn:multi_op_equi}) is actually equivalent to a single objective problem of any one of the four objectives, which can be easily solved by the "fmincon" function in MATLAB.
Since the ANN models are very fast, the computational burden of this method is acceptable in this study.

\begin{subequations}
  \label{eqn:multi_op_equi}
  \begin{equation}
      \min_{Q_{\rm st},Q_{\rm air},T_{\rm fur},V_{\rm cell}} IH_{I},IH_{T},V_{\rm cell},-I_{\rm tot}
  \end{equation}
  \begin{equation}
    {\rm s.t.}\quad T_{\rm fur}= T_{\rm fur,i}, SU= SU_{\rm j}
  \end{equation}
\end{subequations}

\subsection{Selection of the optimal operating condition}
\label{sec:3-4}
How to select the "optimal" solution on the Pareto front is an important problem, because it is impossible to find a solution that dominates the other solutions on all of the objectives.
This paper adopts the linear programming technique for multidimensional analysis of preference (LINMAP) method to make decision \cite{ZHU2021}.

The LINMAP method is based on Euclidean distance.
Denote the $i$th objective of the $j$th solutions on the Pareto front as $f_{ij}$. 
First, this method normalizes the solution by Equation (\ref{eqn:LINMAP_norm}).
$f_{i,{\rm max}}$ and $f_{i,{\rm min}}$ are the maximum and minimum values of the $i$th objective.
Then, the Euclidean distance between the $j$th solution and the ideal solution is calculated by Equation (\ref{eqn:LINMAP_dist}).
$f_{i}^{\rm ideal}$ is the ideal value of the $i$th objective.
It equals to 1 if the $i$th objective is to be maximized, while it equals to 0 if the $i$th objective is to be minimized.
$w_i$ is the weight of the $i$th objective, reflecting the preference for the objective.
Two sets of weights are used, as shown in Table \ref{tab:weightcase}.
It will be discussed later in Section \ref{sec:4-4-2} that it is possible to adjust the performances of the selected optimal solution on different objectives by simply adjusting the weights of different objectives.
Finally, the solutions on the Pareto front are ranked according to $d_{j}$.
The solution with the lowest $d_{j}$ is chosen as the "optimal" solution because it is "nearest" to the ideal solution.
\begin{equation}
  \label{eqn:LINMAP_norm}
  f_{ij}^{\rm norm}=\frac{f_{ij}-f_{i,{\rm min}}}{f_{i,{\rm max}}-f_{i,{\rm min}}}
\end{equation}
\begin{equation}
  \label{eqn:LINMAP_dist}
  d_{j}=\sqrt{\sum_{i=1}^n w_i\left(f_{ij}^{\rm norm}-f_{i}^{\rm ideal}\right)^2}
\end{equation}

\begin{table}[H]
  \footnotesize
  \caption{Weights for different objectives.}
   \centering
   \begin{tabular}{ccccccc}
     \toprule
        &$w_{IH_{I}}$&$w_{IH_{T}}$&$w_{V_{\rm cell}}$&$w_{SU}$&$w_{T_{\rm fur}}$&$w_{I_{\rm tot}}$\\
     \midrule
      {Case 1}&1&1&1&1&1&1\\
      {Case 2}&1&1&1&5&1&1\\
     \bottomrule
   \end{tabular}
   \label{tab:weightcase}
\end{table}

\section{Result and Discussion}
\FloatBarrier
\label{sec:4}
\subsection{Model validation}
\label{sec:4-1}
\begin{figure}[H]
  \centering
  \includegraphics[width=1\textwidth]{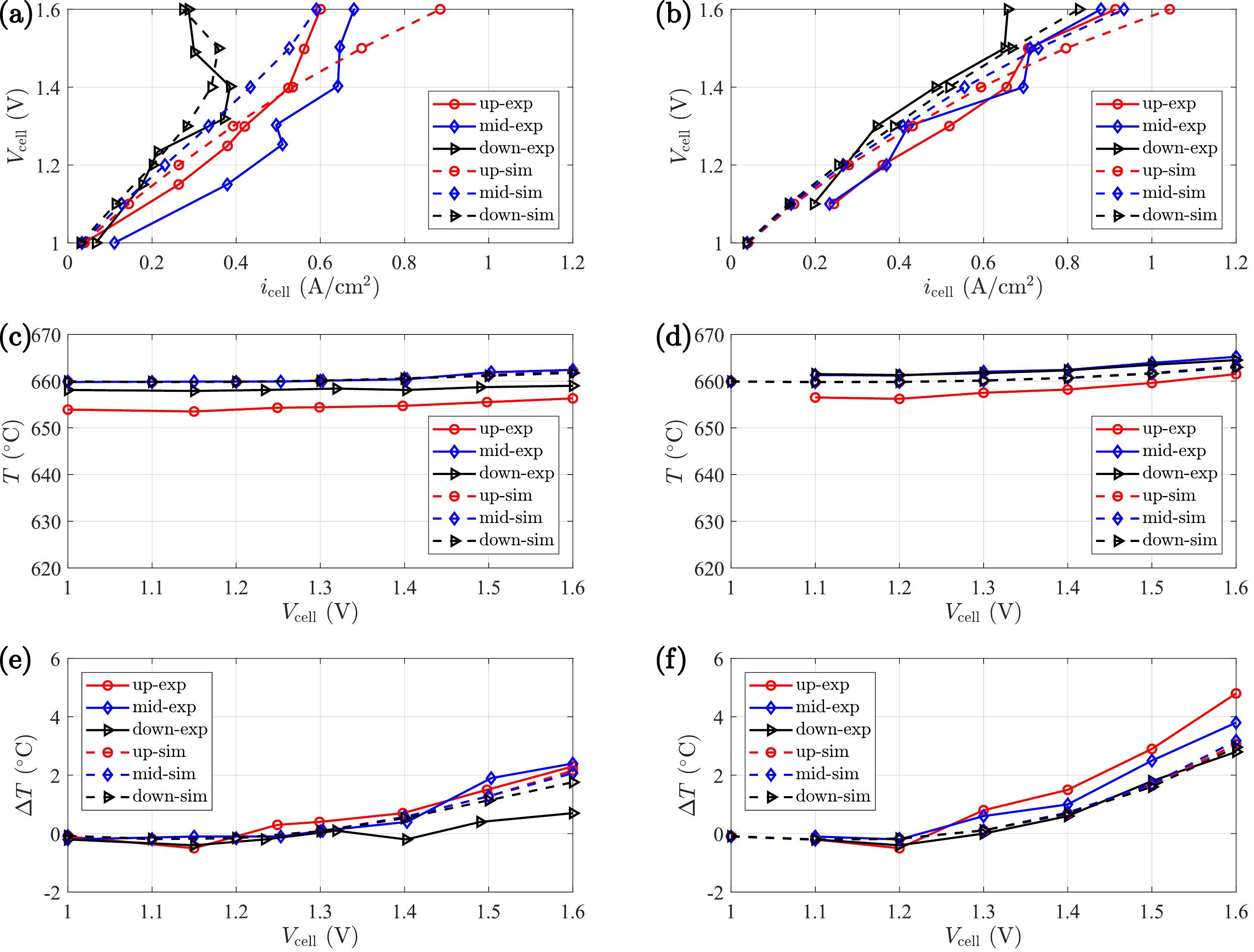}
  \caption{Experimental validation of the 3D multiphysics model. IV curves under (a) Condition 1 and (b) Condition 2. Temperature curves under (c) Condition 1 and (d) Condition 2. Temperature change curves under (e) Condition 1 and (f) Condition 2.}
  \label{fig:ExpVer}
\end{figure}

The accuracy of the 3D model and the ANN surrogate models is verified in this section.
{In order to verify the accuracy of the 3D multiphysics model, it is simulated under the experiment conditions listed in Table \ref{tab:expcondition}.
The simulated IV curves and temperature curves are compared with the experimental results, as shown in Figure \ref{fig:ExpVer}.}
"up", "mid", and "down" represent the up-stream segment, middle-stream segment, and down-stream segment of the SOEC, respectively, as shown in Figure \ref{fig:studyoverview}(d).

{Figure \ref{fig:ExpVer}(a)(b) compare the measured and simulated IV curves of different segments.
The solid lines represent the experimental results, which show that the down-stream segment has the highest voltage due to the consumption of steam along the flow direction.
This is also observed on the simulated IV curves represented by the dashed lines.
In Figure \ref{fig:ExpVer}(a) where the steam flow rate is 40 sccm (Condition 1), the down-stream current shows a decreasing trend after the cell voltage exceeds 1.5 V.
Such a trend is not observed in Figure \ref{fig:ExpVer}(b) where the steam supply is 120 sccm (Condition 2), but mass transport limit can be also observed.
Therefore, the consumption of steam is a critical factor that induces the inhomogeneous current distribution, and high steam utilization is unfavorable for the homogeneity of current.}

{Figure \ref{fig:ExpVer}(c)(d) compare the temperature profiles of experiments and simulations.
It shows that the simulated temperature differences between segments are negligible, while the measured temperatures differences are around \SI{5}{\celsius}.
The measurements show that the up-stream temperature is the lowest and the middle-stream temperature is the highest, which is probably caused by the inhomogeneous temperature distribution inside the furnace.
The middle-stream segment is near to the furnace center and thus has the highest temperature. 
The up-stream segment is far from the furnace center, making the up-stream temperature the lowest.
These factors are not simulated by the 3D model, because the model complexity will be significantly enhanced if the furnace and the inlet tubes are modelled.
In order to validate the model, the temperature rise of each segment against the open circuit state is calculated, as shown in Figure \ref{fig:ExpVer}(e)(f).
In this way, the effect of the inhomogeneous furnace temperature distribution is partly eliminated.
Figure \ref{fig:ExpVer}(e)(f) show that the simulated temperature rises match well with experimental results, indicating that the heat transfer model and boundary conditions are appropriate.}

The errors between experiments and simulations are analyzed as follows.
\begin{enumerate}
  \item Figure \ref{fig:ExpVer}(a)(b) shows that the measured voltage of the middle-stream segment is lower than that of the up-stream segment, conflicting with the steam concentration distribution.
  The inhomogeneous furnace temperature is a possible reason.
  The high temperature of the middle-stream segment can make it more active than the up-stream segment.
  Different contact resistances of different segments can also cause errors.
  \item It is observed that the current was unstable during potentiostatic measurements in the experiments, which is probably caused by local electrode re-oxidation or steam condensation in the inlet tube and the outlet tube.
\end{enumerate}
{Generally speaking, the trends of the current density distribution and the temperature distribution are predicted correctly by the 3D model.
Therefore, it is considered that the 3D model can reflect the actual behavior of a planar SOEC cell and thus is reliable for verifying the optimization method proposed in Section \ref{sec:3}.}

\begin{figure}[H]
  \centering
  \includegraphics[width=1\textwidth]{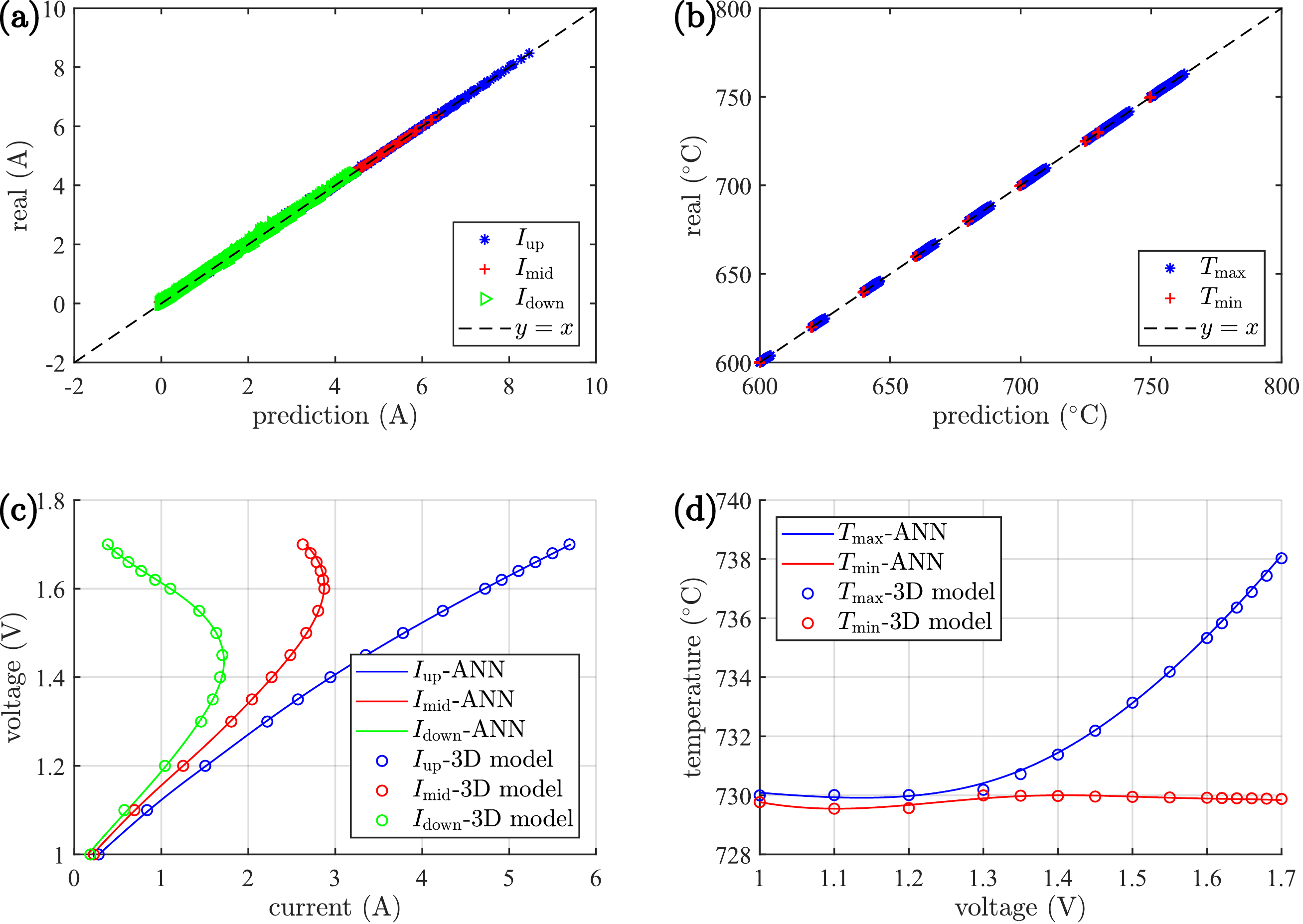}
  \caption{Fitting results of the ANN models. (a) Current distribution. (b) Temperature distribution. (c) Verification of current distribution and (d) temperature distribution when $T_{\rm fur}=730$ \SI{}{\celsius}, $Q_{\rm air}=100$ sccm, and $Q_{\rm st}=60$ sccm.}
  \label{fig:NNvali}
\end{figure}

Using the simulation data produced by the 3D model, the ANN surrogate models are trained to predict the current distribution and temperature distribution.
The accuracy of the ANN models on the whole dataset is shown in Figure \ref{fig:NNvali}(a)(b).
The horizontal axes and the vertical axes represent the prediction of the ANN models and the simulation results of the 3D model.
The closer the data point is to the dashed line, the more accurate the ANN models are.
Figure \ref{fig:NNvali}(a)(b) show that both the current distribution and the temperature distribution are predicted accurately by the ANN surrogate models.
Figure \ref{fig:NNvali}(c)(d) demonstrate the IV curves and temperature curves predicted by the ANN models when $T_{\rm fur}=730$ \SI{}{\celsius}, $Q_{\rm air}=100$ sccm, and $Q_{\rm st}=60$ sccm.
It is found that the ANN models can accurately reproduce the IV curves and temperature curves simulated with the 3D model.
The ANN models take far less computational cost than the original 3D model, which is critical for conducting the optimization study.
On the contrary, it takes approximately 10 min for the 3D model to simulate a single stationary case, making it unsuitable to be integrated with optimization problems.

\FloatBarrier
\subsection{Influencing factors of inhomogeneity}
\FloatBarrier
\label{sec:4-2}
Before solving the optimization problem, this section studies how the operating conditions influence different objectives.
The Sobol sensitivities of the performance indices defined in Equations (\ref{eqn:unindex}) against the inputs are calculated as introduced in Section \ref{sec:3-1}.
The results are presented in Table \ref{tab:SobolSensitivity}.
Table \ref{tab:SobolSensitivity} indicates that, within the input ranges, $Q_{\rm air}$ has little influence on the performance indices because the sensitivities corresponding to $Q_{\rm air}$ are close to zero.
For industrial SOEC systems, air is used for temperature regulation and blowing out the produced oxygen.
However, in the 3D model used by this study, the effect of air cooling is insignificant compared with the heat exchange between the furnace and the cell assembly, because the surface area of the cell assembly is large.
Since $Q_{\rm air}$ barely influences the simulation results, it is fixed at 100 sccm during the following analysis, and only three decision variables, $T_{\rm fur}$, $V_{\rm cell}$, and $Q_{\rm st}$, remain in the optimization.
\begin{table}[H]
  \footnotesize
  \caption{{Sobol sensitivity.}}
   \centering
   \begin{tabular}{cccccc}
     \toprule
      && $T_{\rm fur}$ & $Q_{\rm air}$ & $Q_{\rm st}$& $V_{\rm cell}$\\
      \hline
     \multirow{2}{*}{$SU$}&$S_i$&0.0698&9.0065e-6&0.7880&0.1015\\
     &$ST_{i}$&0.0872&1.2445e-4&0.8277&0.1267\\
     \multirow{2}{*}{$IH_{I}$}&$S_i$&0.3625&0.0010&0.1547&0.4267\\
     &$ST_{i}$&0.4047&9.9666e-4&0.1847&0.4724\\
     \multirow{2}{*}{$IH_{T}$}&$S_i$&0.0438&-4.1308e-5&0.0037&0.9437\\
     &$ST_{i}$&0.0570&9.1617e-5&0.0060&0.9580\\
     \bottomrule
   \end{tabular}
   \label{tab:SobolSensitivity}
\end{table}

\begin{figure}[H]
  \centering
  \includegraphics[width=1\textwidth]{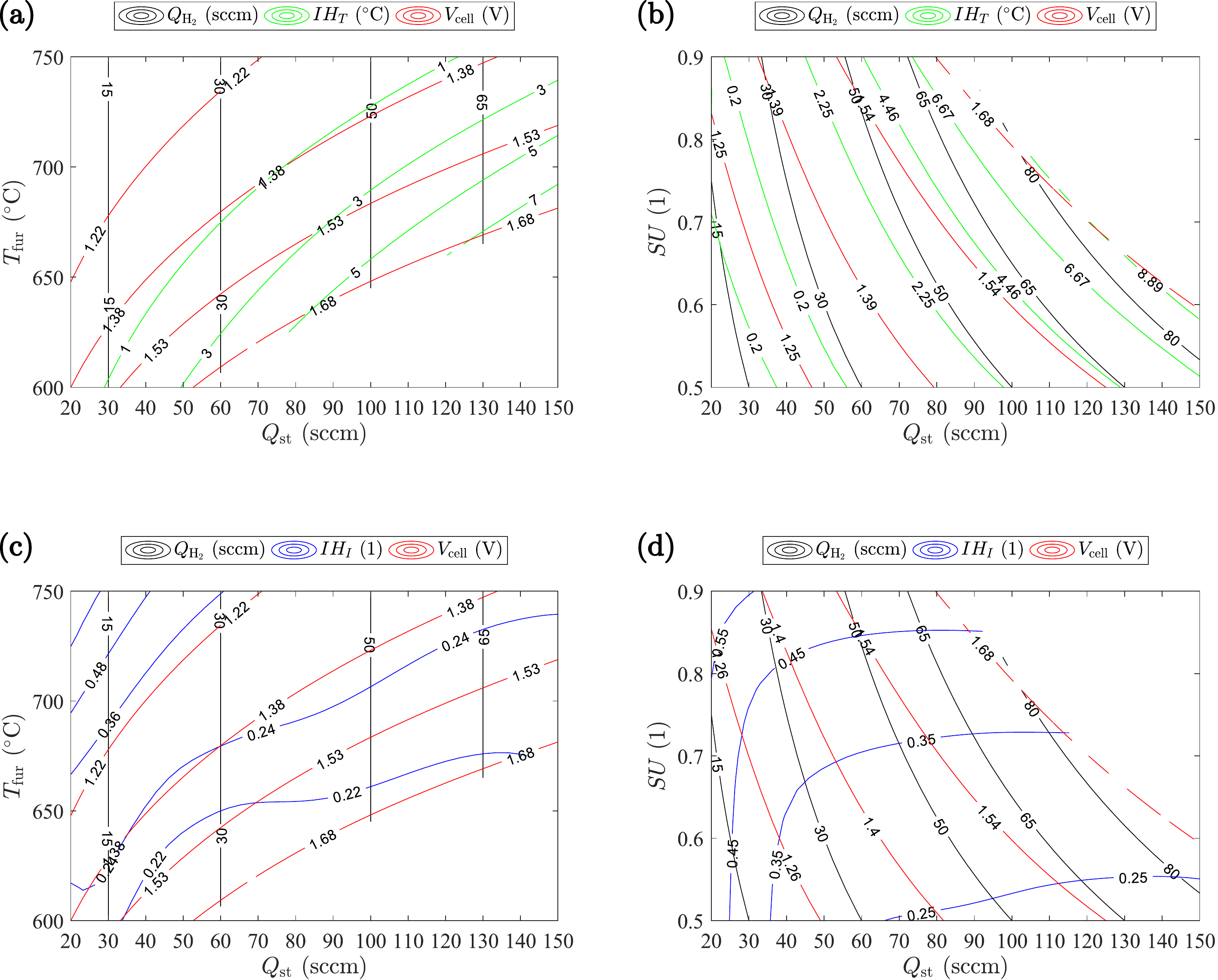}
  \caption{2D projection of $IH_{T}^{*}(T_{\rm fur},Q_{\rm st},SU)$, $V_{\rm cell}^{*}(T_{\rm fur},Q_{\rm st},SU)$, and $Q_{\rm H_2}^{*}(T_{\rm fur},Q_{\rm st},SU)$ on (a) $SU=0.5$ and (b) $T_{\rm fur}=700^\circ{\rm C}$. 2D projection of $IH_{I}^{*}(T_{\rm fur},Q_{\rm st},SU)$, $V_{\rm cell}^{*}(T_{\rm fur},Q_{\rm st},SU)$, and $Q_{\rm H_2}^{*}(T_{\rm fur},Q_{\rm st},SU)$ on (c) $SU=0.5$ and (d) $T_{\rm fur}=700^\circ{\rm C}$. For a better visualization effect, the contour lines are not plotted equidistantly.}
  \label{fig:Single-result}
\end{figure}

{A three dimensional grid of $T_{\rm fur}$, $Q_{\rm st}$, and $SU$ is created.
At each grid point, the surrogate models are used to solve for $V_{\rm cell}$ in order to reach the specified $SU$ corresponding to the grid point.
The "fmincon" function in MATLAB is used to search for the solution.
The solutions are denoted as $V_{\rm cell}^{*}(T_{\rm fur},Q_{\rm st},SU)$, and the corresponding performance indices can be denoted as $IH_{T}^{*}(T_{\rm fur},Q_{\rm st},SU)$, $IH_{I}^{*}(T_{\rm fur},Q_{\rm st},SU)$, and $Q_{\rm H_2}^{*}(T_{\rm fur},Q_{\rm st},SU)$.
They can be viewed as multivariate functions defined in a three-dimensional input space.
Their values on the 2D planes that represent $SU=0.5$ and $T_{\rm fur}=700^\circ{\rm C}$ are illustrated in Figure \ref{fig:Single-result}.
For clearness, $IH_{T}$ and $IH_{I}$ are plotted in separate.
}

{Figure \ref{fig:Single-result}(a)(b) show $Q_{\rm H_2}^*$, $IH_{T}^*$, and $V_{\rm cell}^*$.
The following analysis is based on a fixed hydrogen production $Q_{\rm H_2}$, which is equivalent to a fixed total current $I_{\rm tot}$.
Take the contour lines of $Q_{\rm H_2}=50$ sccm for example.
Figure \ref{fig:Single-result}(a), where $SU$ is fixed at 0.5, shows that both $V_{\rm cell}$ and $IH_{T}$ decreases as $T_{\rm fur}$ increases on the contour line of $Q_{\rm H_2}=50$ sccm.
The reason is that increasing $T_{\rm fur}$ can lower down the overpotentials and thus lower down $V_{\rm cell}$, which further lower down $IH_{T}$ due to the reduced heat production.
Figure \ref{fig:Single-result}(b), where $T_{\rm fur}$ is fixed at \SI{700}{\celsius}, shows that both $V_{\rm cell}$ and $IH_{T}$ increases as $SU$ increases, due to the increased concentration overpotentials.
Increasing $SU$, decreasing $T_{\rm fur}$, and decreasing $V_{\rm cell}$ are desired for system efficiency, while decreasing $IH_{I}$ and $IH_{T}$ is desired for durability.
Therefore, the abovementioned relationships are summarized as follows.\\
(\#1) Decreasing $T_{\rm fur}$ conflicts with decreasing $V_{\rm cell}$.\\
(\#2) Decreasing $T_{\rm fur}$ conflicts with decreasing $IH_{T}$.\\
(\#3) Increasing $SU$ conflicts with decreasing $V_{\rm cell}$.\\
(\#4) Increasing $SU$ conflicts with decreasing $IH_{T}$.\\
(\#5) From (\#1)-(\#4), it is inferred that decreasing $V_{\rm cell}$ accords with decreasing $IH_{T}$.}

{Figure \ref{fig:Single-result}(c)(d) illustrate $Q_{\rm H_2}^*$, $IH_{I}^*$, and $V_{\rm cell}^*$, among which $Q_{\rm H_2}^*$ and $V_{\rm cell}^*$ are the same with Figure \ref{fig:Single-result}(a)(b).
Also take the contour lines of $Q_{\rm H_2}=50$ sccm for example.
Figure \ref{fig:Single-result}(c)(d) show that $IH_{I}$ increases when $T_{\rm fur}$ and $SU$ increases, respectively.
The latter phenomenon is caused by the enhanced inhomogeneity of steam concentration when $SU$ increases, while the former phenomenon is explained qualitatively in \ref{sec:appendix1}
These relationships are summarized as follows.\\
(\#6) Decreasing $T_{\rm fur}$ accords with decreasing $IH_{I}$.\\
(\#7) Increasing $SU$ conflicts with decreasing $IH_{I}$.\\
}

\begin{table}
  \footnotesize
  \caption{Conflicting relationships between multiple objectives at fixed $Q_{\rm H_2}$, i.e., fixed $I_{\rm tot}$.}
   \centering
   \begin{tabular}{cc|ccc|cc}
     \toprule
     &&\multicolumn{3}{c|}{Efficiency}&\multicolumn{2}{c}{Durability}\\
     && \makecell[c]{Increasing\\$SU$}& \makecell[c]{Decreasing\\ $T_{\rm fur}$} & \makecell[c]{Decreasing\\ $V_{\rm cell}$} & \makecell[c]{Decreasing\\ $IH_{I}$}& \makecell[c]{Decreasing\\ $IH_{T}$}  \\
     \midrule
    \multirow{3}{*}{\makecell{\\ \\Efficiency}}&\makecell[c]{Increasing\\$SU$}&-&independent&conflict (\#3)&conflict (\#7) &conflict(\#4)\\ 
    &\makecell[c]{Decreasing\\ $T_{\rm fur}$}&-&-&conflict (\#1)&accord (\#6)&conflict (\#2)\\
    &\makecell[c]{Decreasing\\ $V_{\rm cell}$}&-&-&-&non-monotonic&accord (\#5)\\
    \hline
    \multirow{2}{*}{\makecell[c]{\\Durability}}&\makecell[c]{Decreasing\\ $IH_{I}$}&-&-&-&-&non-monotonic\\
    &\makecell[c]{Decreasing\\ $IH_{T}$}&-&-&-&-&-\\
     \bottomrule
   \end{tabular}
   \label{tab:Contradicting}
\end{table}

The abovementioned relationships between multiple objectives at a fixed $Q_{\rm H_2}$ are summarized in Table \ref{tab:Contradicting}.
Only half of the table is filled due to the symmetricity.
In this table, 
'non-monotonic' means that the relationship is non-monotonic, 
and 'independent' means that $T_{\rm fur}$ and $SU$ are two variables that are varied independently in the analysis.
{This table conveys information as follows.
\begin{itemize}
  \item The factors affecting system efficiency, including $SU$, $T_{\rm fur}$, and $V_{\rm cell}$, conflict with each other.
   An optimal trade-off among these factors should be searched to maximize the system efficiency.
  This problem can be handled by system-level optimization studies when the system structure and parameters are determined.
  For example, Xing et al. \cite{XING2018} optimize the steam utilization, operating temperature, and current to maximize the system efficiency under different powers.
  \item There are complex relationships between the factors affecting efficiency and durability. 
  The relationship between $IH_{I}$ and $IH_{T}$, both of which affect durability, is also complex.
  However, system-level studies considering these factors have not been reported to the authors' best knowledge, because the model used by current system-level optimization studies cannot simulate the inhomogeneity accurately, as analyzed in Section \ref{sec:1}.
\end{itemize}
}
The following section conducts stack-level optimization to balance efficiency and homogeneity.
Although the system efficiency is not calculated, stack manufacturers can parameterize the relationships shown in Figure \ref{fig:Single-result} by curve fitting, and then deliver the results to system operators to conduct system-level optimizations.
In this way, the stack behaviors can be modelled more accurately in system-level optimization, and the inhomogeneity can also be considered.

\FloatBarrier
\subsection{Determination of the optimal conditions}
\label{sec:4-4}
\subsubsection{Pareto front under a fixed power and the optimal solution chosen by LINMAP}
\label{sec:4-4-1}
\FloatBarrier

This section solves the multi-objective optimization problem, Equation (\ref{eqn:multi_op_equi}), by grid search and reveal the conflicting relationships between the objectives.
This section will show that the solutions of Equation (\ref{eqn:multi_op_equi}) form the solutions of the original problem, Equation (\ref{eqn:multi_op}).

Table \ref{tab:Contradicting} is inspiring for solving the multi-objective optimization problem.
It reveals that decreasing $IH_{T}$ accords with decreasing $V_{\rm cell}$ when $I_{\rm tot}$ is fixed.
Meanwhile, with a fixed total power $P_{\rm ele}$ in Equation (\ref{eqn:multi_op_equi}), increasing $I_{\rm tot}$ accords with decreasing $V_{\rm cell}$.
Moreover, Figure \ref{fig:Single-result}(c) shows that increasing $Q_{\rm st}$ will lower down $IH_{I}$ when $SU$ is fixed, which indicates that increasing $I_{\rm tot}$ accords with decreasing $IH_{I}$.
In other words, the four objectives in Equation (\ref{eqn:multi_op_equi}) accord with each other, and Equation (\ref{eqn:multi_op_equi}), a multi-objective optimization problem, is actually equivalent to a single objective problem of any one of the four objectives.
The "fmincon" function in MATLAB is used to search for the solutions.

\begin{figure}
  \centering
  \includegraphics[width=1\textwidth]{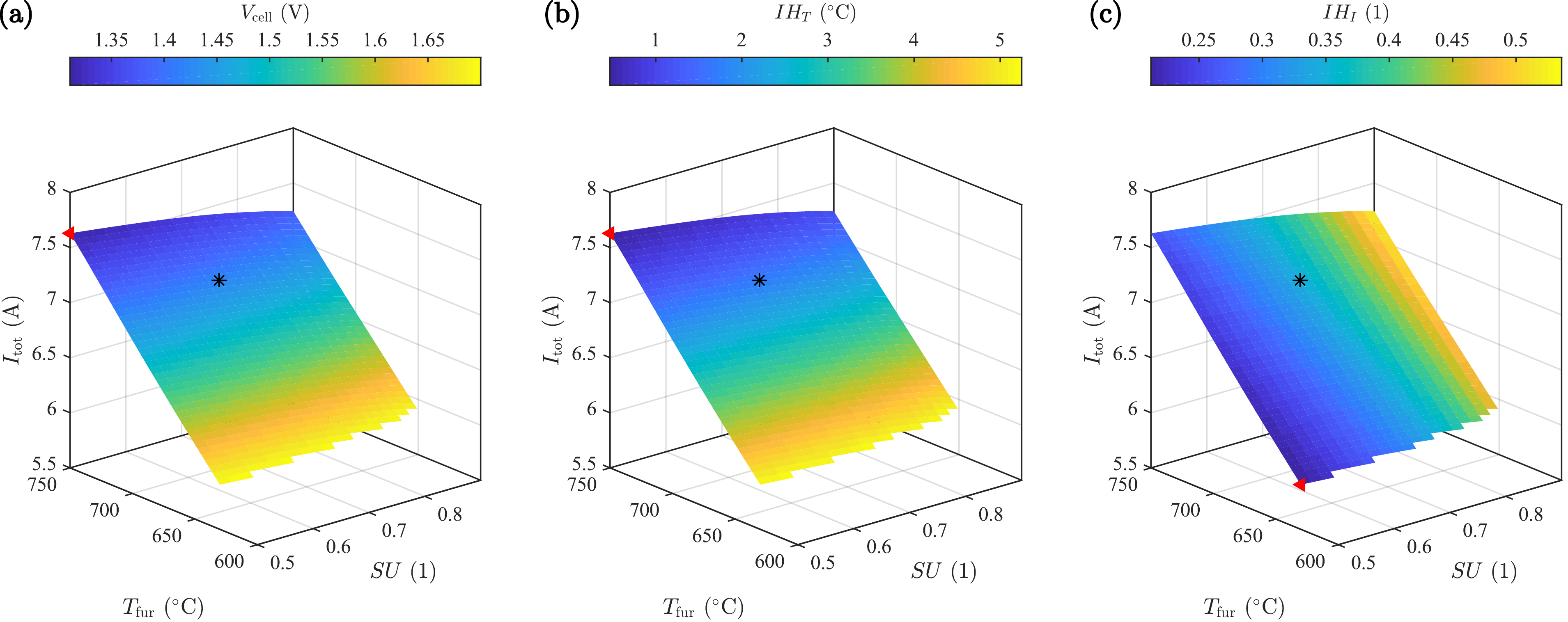}
  \caption{Solutions of Equation (\ref{eqn:multi_op_equi}) under different $T_{\rm fur}$ and $SU$ when $P_{\rm ele}=10$ W. The color bars represent (a) $V_{\rm cell}$, (b) $IH_{T}$, and (c) $IH_{I}$. The red triangles represent the best values of $V_{\rm cell}$, $IH_{T}$, and $IH_{I}$, while the black asterisks represent the optimal solution chosen by the LINMAP method.}
  \label{fig:map_NUI_NUT}
\end{figure}

The optimization result when $P_{\rm ele}=10$ W is illustrated in Figure \ref{fig:map_NUI_NUT}.
The color in Figure \ref{fig:map_NUI_NUT}(a)-(c) represents $V_{\rm cell}$, $IH_{T}$, and $IH_{I}$, respectively.
Figure \ref{fig:map_NUI_NUT}(a) shows that a high $T_{\rm fur}$ and a low $SU$ are preferred to minimize $V_{\rm cell}$ and thus maximize $I_{\rm tot}$, because the ohmic and activation overpotentials both decrease as $T_{\rm fur}$ increases and $SU$ decreases.
Since the decreased overpotentials lower down the heat production,  $IH_{T}$ also reaches its minimum when $T_{\rm fur}$ is maximized and $SU$ is minimized, as shown in Figure \ref{fig:map_NUI_NUT}(b).
Figure \ref{fig:map_NUI_NUT}(c) shows that a low $SU$ is preferred to minimize $IH_{I}$, because the concentration gradient along the flow direction decreases as $SU$ decreases.
$SU$ is barely influenced by $T_{\rm fur}$.
The best values of $V_{\rm cell}$, $IH_{T}$, and $IH_{I}$ are marked with red triangles in Figure \ref{fig:map_NUI_NUT}(a)-(c).
Additionally, a high $SU$ and a low $T_{\rm fur}$ is generally preferred for an SOEC system to reduce the power consumption of the BOPs and thus improve the system efficiency.
According to the analysis, for any point on the surfaces in Figure \ref{fig:map_NUI_NUT}(a)-(c), it is impossible to improve one of the objectives without satisfying any of the rest objectives. 
Therefore, these surfaces form the Pareto front of the original multi-objective problems Equation (\ref{eqn:multi_op}) when the electrolysis power $P_{\rm ele}$ is fixed at 10 W.

The best value of each individual objective cannot be reached simultaneously.
To enable decision making on the Pareto front, the LINMAP method introduced in Section \ref{sec:3-3} is used to choose the "optimal" solution on the Pareto front.
The weights in Equation (\ref{eqn:LINMAP_dist}) are set as Case 1, as listed in Table \ref{tab:weightcase}, which means that all the objectives are treated equally.
The selected optimal solution is marked by black asterisks in Figure \ref{fig:map_NUI_NUT}. 

A comparison between the optimal solution and the best/worst value of each individual objective on the Pareto front is listed in Table \ref{tab:SolCompare}.
The relative distance between the optimal solution and the best/worst value of each objective is calculated, represented by percentiles.
Table \ref{tab:SolCompare} shows that the relative distances to the best values of $IH_{I}$, $IH_{T}$, $V_{\rm cell}$, and $I_{\rm cell}$ are less than 50\%, while the relative distances to the best values of $T_{\rm fur}$ and $SU$ exceed 50\%.
Therefore, the optimal solution chosen by the LINMAP method perform better on $IH_{I}$, $IH_{T}$, $V_{\rm cell}$, and $I_{\rm cell}$ than on $SU$ and $T_{\rm fur}$.
It indicates that the selected optimal solution tends to sacrifice $SU$ and $T_{\rm fur}$ to improve the other four objectives.
In other words, the optimal solution tends to choose a high $T_{\rm fur}$ and a low $SU$, which degrade system efficiency due to increased energy consumption, to obtain a low $V_{\rm cell}$, a high $I_{\rm cell}$, and a low inhomogeneity.
The reason is that $IH_{I}$, $IH_{T}$, $V_{\rm cell}$, and $I_{\rm cell}$ accord with each other as discussed in the beginning of this section.
Therefore, it is beneficial to sacrifice $T_{\rm fur}$ and $SU$ slightly to improve the other four objectives.
Meanwhile, a low $V_{\rm cell}$ and a high $I_{\rm cell}$ are beneficial for system efficiency, which can partly balance out the efficiency reduction caused by a high $T_{\rm fur}$ and a low $SU$.
The results show that the LINMAP method can effectively achieve a trade-off among the multiple conflicting objectives.
\begin{table}
  \footnotesize
  \caption{Comparison between the solution determined by the LINMAP method and the best value of each individual objective when $P_{\rm ele}=10$ W.}
   \centering
   \begin{tabular}{cccccccc}
     \toprule
        &&$IH_{I}$&$IH_{T}$&$V_{\rm cell}$&$SU$&$T_{\rm fur}$&$I_{\rm tot}$\\
     \midrule
      \multicolumn{2}{c}{Best value}&0.21&0.47&1.31&0.90&630&7.63\\
      \multicolumn{2}{c}{Worst value}&0.54&5.25&1.70&0.50&750&5.88\\
      \hline
      \multirow{3}{*}{\makecell[c]{LINMAP\\ decision}}&Value&0.35&1.69&1.42&0.7&720&7.05\\
      &\makecell[c]{Relative distance to the best value}&43.2\%&25.5\%&27.7\%&50.0\%&75.0\%&33.2\%\\
      &\makecell[c]{Relative distance to the worst value}&56.8\%&74.5\%&72.3\%&50.0\%&25.0\%&66.8\%\\
     \bottomrule
   \end{tabular}
   \label{tab:SolCompare}
\end{table}

\subsubsection{Pareto fronts under different powers and the optimal operating curves}
\label{sec:4-4-2}
\begin{figure}[H]
  \centering
  \includegraphics[width=1\textwidth]{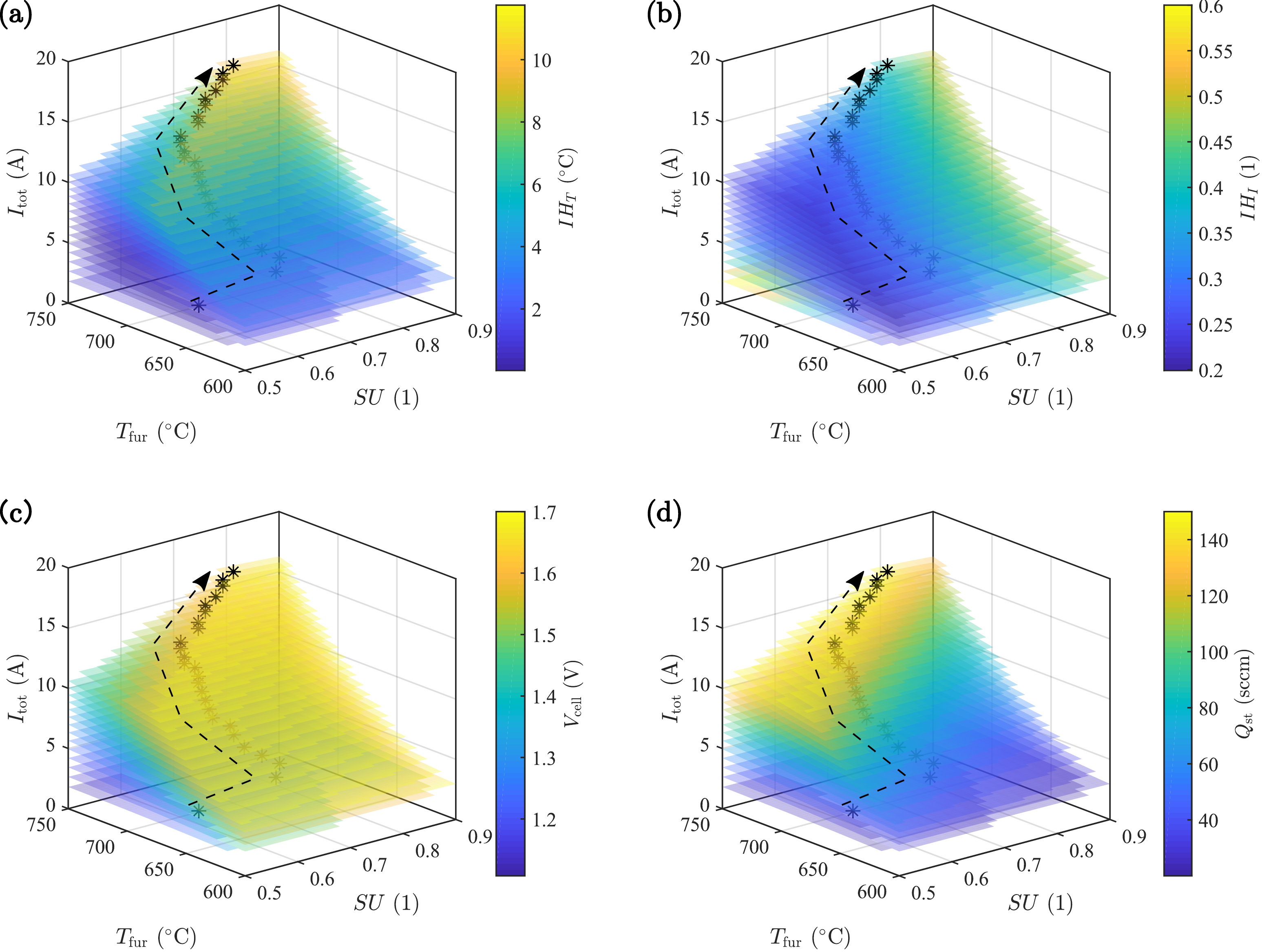}
  \caption{Pareto fronts when $P_{\rm ele}$ increases from 2 W to 27 W with an interval of 1 W. The optimal solutions chosen by the LINMAP method at different $P_{\rm ele}$ are marked with black asterisks. The arrows indicate the directions in which the optimal solutions change as $P_{\rm ele}$ increases.}
  \label{fig:map_NUI_V}
\end{figure}

When the target electrolysis power $P_{\rm ele}$ changes, the corresponding Pareto front moves. 
The Pareto fronts when $P_{\rm ele}$ increases from 2 W to 27 W (with an interval of 1 W) are shown in Figure \ref{fig:map_NUI_V}, represented by the surfaces.
In Figure \ref{fig:map_NUI_V}(a)-(d), the color of the surfaces represents $IH_{T}$, $IH_{I}$, $V_{\rm cell}$, and $Q_{\rm st}$, respectively.
The heights of these surfaces indicate that the hydrogen production (i.e., $I_{\rm tot}$) generally rises as $P_{\rm ele}$ increases.

Figure \ref{fig:map_NUI_V}(a)(c) show that both $IH_{T}$ and $V_{\rm cell}$ rise when $P_{\rm ele}$ increases, due to the increased current.
Therefore, for the operation of an SOEC system, it is necessary to control the voltage to avoid a large temperature gradient inside the stack.
Combining Figure \ref{fig:map_NUI_NUT}(c) and Figure \ref{fig:map_NUI_V}(b), it is found that $IH_{I}$ is significantly influenced by $SU$.
Therefore, it is necessary to control $SU$ and ensure a homogeneous distribution of steam between different cells and channels to improve the homogeneity of current distribution.
The conflicting relationship between $SU$ and $IH_{I}$, demonstrated in Figure \ref{fig:map_NUI_V}(b), can be parameterized and delivered to the system operator, so that they can choose a proper $SU$ to balance efficiency and homogeneity.
Additionally, it is found that $IH_{I}$ becomes high at the	bottom left corner in Figure \ref{fig:map_NUI_V}(b), indicating that a high temperature and a low steam utilization are unfavorable when $P_{\rm ele}$ is low.
From Figure \ref{fig:map_NUI_V} (c)(d), it is found that the feasible ranges of $SU$ and $T_{\rm fur}$ are limited when $P_{\rm ele}$ is too low or too high, limited by the ranges of $Q_{\rm st}$ ([20,150] sccm) and $V_{\rm cell}$ ([1.0,1.7] V).
At a low $P_{\rm ele}$, e.g. 2 W, a high $SU$ cannot be achieved when $Q_{\rm st}$ reaches its lower bound of 20 sccm, because increasing $SU$ will increase the current and overpotential, making $P_{\rm ele}$ exceed 2 W.
The electrolysis power $P_{\rm ele}$ cannot reach zero in this study because of the lower bounds of $Q_{\rm st}$ and $SU$ (20 sccm and 0.5, respectively).
On the contrary, at a high $P_{\rm ele}$, e.g. 27 W, $SU$ cannot be further lowered down when $Q_{\rm st}$ reaches its upper bound of 150 sccm, because the current and overpotential will decrease and the cell cannot reach the target power.
Also, a high $T_{\rm fur}$ is necessary at a high $P_{\rm ele}$ because lowering down $T_{\rm fur}$ will result in $V_{\rm cell}$ exceeding its upper bound of 1.7 V.

\begin{figure}[H]
  \centering
  \includegraphics[width=1\textwidth]{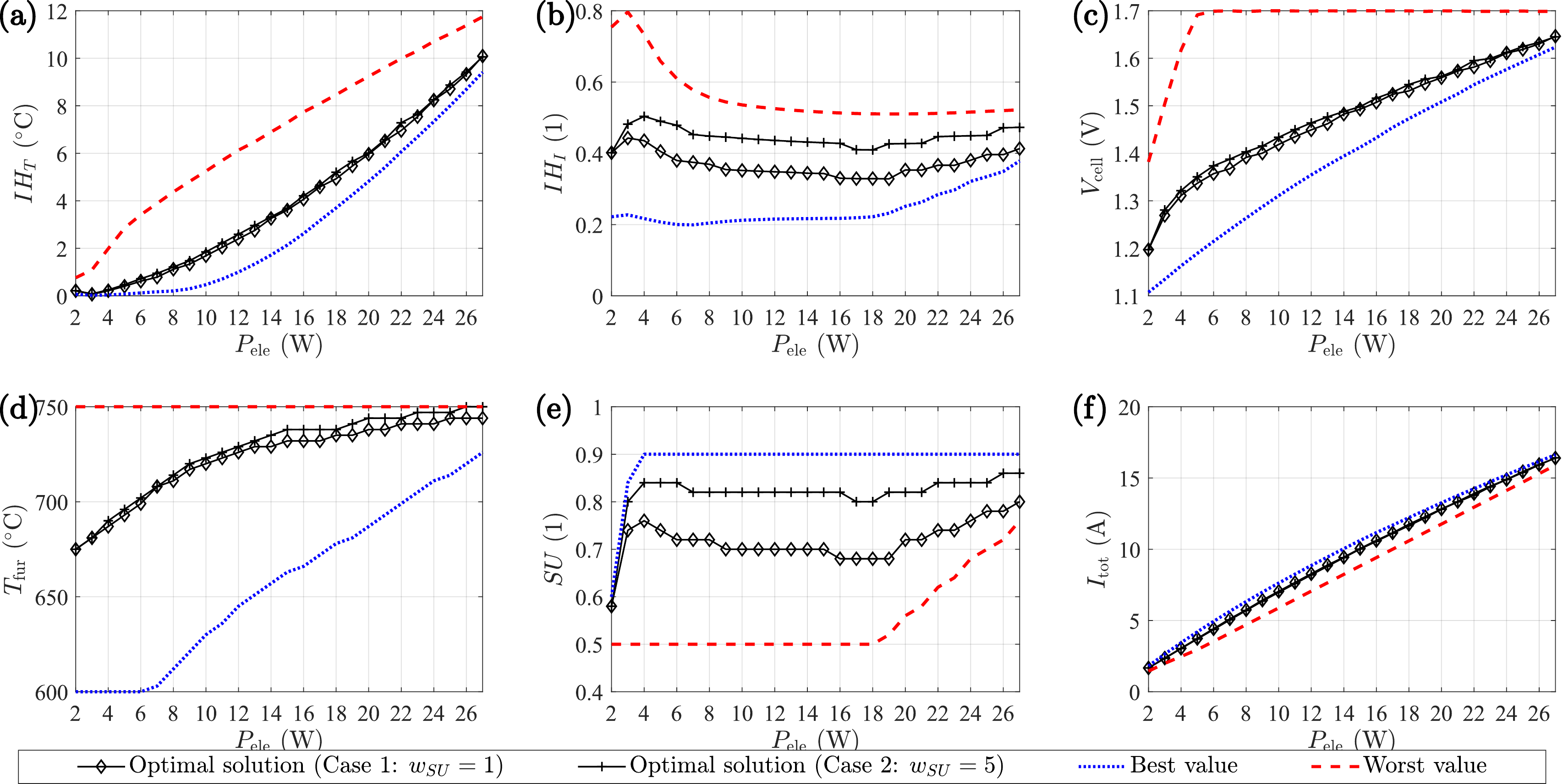}
  \caption{Optimal operating curves determined by the LINMAP method. (a)-(f) show the values of $IH_{T}$, $IH_{I}$, $V_{\rm cell}$, $T_{\rm fur}$, $SU$, and $I_{\rm tot}$ corresponding to the optimal solutions.}
  \label{fig:opt_curve}
\end{figure}

Figure \ref{fig:map_NUI_V} also illustrates the optimal solutions chosen by the LINMAP method with all the weights equal to 1 (Case 1 in Table \ref{tab:weightcase}), marked by black asterisks.
The dashed black arrows represent the direction in which the optimal solutions change as $P_{\rm ele}$ increases.
The values of these optimal solutions are presented in Figure \ref{fig:opt_curve}, represented by the black lines with diamond markers.
The best and worst values of the six objectives are also illustrated in Figure \ref{fig:opt_curve}, represented by the blue and red lines, respectively.

Figure \ref{fig:opt_curve}(a)(c)(f) show that $IH_{T}$, $V_{\rm cell}$, and $I_{\rm tot}$ are all close to their best values (blue lines) under different $P_{\rm ele}$, and they rise monotonically when $P_{\rm ele}$ increases.
Figure \ref{fig:opt_curve}(b)(e) show that the trends of $IH_{I}$ and $SU$ are complex when $P_{\rm ele}$ increases.
When $P_{\rm ele}$ increases from 2 W to 4 W, $SU$ rises from c.a. 0.58 to 0.75 (Case 1).
When $P_{\rm ele}$ increases from 4 W to 18W, $SU$ drops slowly to c.a. 0.69.
As $P_{\rm ele}$ further increases, $SU$ increases to c.a. 0.8.
Correspondingly, $NU_I$ first increases from c.a. 0.4 to 0.43, then drops slowly to c.a. 0.35, and finally rises to c.a. 0.4. 
Such trends are caused by the decision procedure of the LINMAP method. 
As introduced in Section \ref{sec:3-4}, the LINMAP method ranks all the solutions on the Pareto fronts by their weighted Euclidean distances to the ideal/best value of each objective.
As $P_{\rm ele}$ increases, the best values and the worst values of all the objectives change and thus the optimal solution chosen by the LINMAP method also changes.
Ignoring the parts of $P_{\rm ele}<4$ W and $P_{\rm ele}>18$ W, the optimal $SU$ is between c.a. 0.7 and 0.75, and the optimal $IH_I$ is between c.a. 0.35 and 0.43 for Case 1. 

To investigate the effect of the weights used by the LINMAP method, the optimal curves under the two cases listed in Table \ref{tab:weightcase} are compared in Figure \ref{fig:opt_curve}.
In Case 1, all the weights are set as 1, indicating that the six objectives are treated equally.
In Case 2, the weight of $SU$ is set as 5, meaning that the preference for $SU$ is higher than those for the other objectives.

Figure \ref{fig:opt_curve}(e) demonstrates that increasing $w_{SU}$ can effectively make the LINMAP method choose optimal solutions that perform better on $SU$.
Within the interval of 4 W$<P_{\rm ele}<18$ W, the optimal $SU$ remains above 0.8 in Case 2.
However, one or some of the other objectives must be sacrificed when the preferred objectives are improved because the essential conflicting relationships between different objectives remain unchanged.
As $SU$ improves, $IH_{I}$ becomes worse/larger.
Compared with Case 1, $IH_{I}$ increases by c.a. 0.1 for each $P_{\rm ele}>4$ W in Case 2.
Figure \ref{fig:opt_curve}(a)(c)(d)(f) show that $IH_{T}$, $V_{\rm cell}$, $T_{\rm fur}$ and $I_{\rm tot}$ are barely influenced because $SU$ is only closely related to $IH_{I}$ in this study.
Therefore, the LINMAP method allows one to obtain customized optimal solutions with improved performances on certain preferred objectives by simply adjusting the weights of different objectives.
Meanwhile, one must remember that such improvement on the preferred objectives comes at the price of some other objectives being sacrificed.

\section{Conclusions}
Segmented SOEC experiments show that the inhomogeneity of current is enhanced under a high steam utilization, indicating that efficiency conflicts with homogeneity.
However, existing operation optimization studies rarely consider current homogeneity, because the models used by them cannot simulate the inhomogeneous distributions of current and temperature reliably.
To handle this problem, this study proposes to combine segmented SOEC experiments, multiphysics simulation, and artificial intelligence, and thus enables the joint optimization of efficiency and homogeneity for an SOEC.

To demonstrate the method, experiments are first conducted on a segmented cathode-supported SOEC.
The current and temperature distributions under two operating conditions are measured.
The experimental data are used to validate the current and temperature distributions predicted by the 3D multiphysics model of the segmented SOEC.
Then, to avoid the high computational cost of the 3D model, ANN surrogate models are trained using the simulation data of the multiphysics model.
The results show that the ANN models match with the multiphysics simulation results accurately.
The ANN surrogate models can predict the SOEC behaviors with low computational cost, which is beneficial for the numerical solution of optimization problems.

With the surrogate models, a multi-objective optimization problem is constructed to optimize the current inhomogeneity, the temperature inhomogeneity, the cell voltage, the steam utilization, the temperature, and the hydrogen production jointly.
These objectives are closely related with system efficiency and durability.
The multi-objective optimization problem is decomposed into a series of single-objective optimization problems.
Their solutions form the Pareto front reflecting the conflicting relationships between these objectives under a fixed electrolysis power.
In this study, the Pareto front shows that decreasing the inhomogeneity accords with decreasing the cell voltage and increasing the hydrogen production, but conflicts with increasing the steam utilization and decreasing the temperature.

To achieve a balance among multiple objectives, the LINMAP method is used to choose the optimal solutions from the Pareto fronts under different electrolysis powers.
The optimal solutions chosen by the LINMAP method indicate a steam utilization of c.a. 0.7.
Under this situation, the down-stream current is 60\%-65\% of the up-stream current. 
Moreover, it is possible to adjust the optimal results according to one's preference by adjusting the weights of different objectives used in the LINMAP method.
In this study, when the weight of steam utilization increase from 1 to 5, the optimal steam utilization rises from 0.7 to 0.8.
As a price, the current inhomogeneity is enhanced.
The down-stream current drops to 50\%-60\% of the up-stream current.
This result indicates that the steam utilization is strongly related to the current inhomogeneity, which should be considered quantitatively in system-level optimizations to balance system efficiency and homogeneity.

The Pareto fronts and the optimal operating curves produced by the proposed method could be easily delivered to the system operators in product datasheets by stack manufacturers.
In this way, system operators can parameterize the Pareto fronts by methods like curve fitting and integrate it into system-level optimization studies, so that they can choose proper operating points to achieve a balance between system efficiency and homogeneity.
In this way, the cooperation between the stack manufacturers and the system operator is enhanced.

\section*{Data availability}
The data produced by the 3D multiphysics model for training and testing the ANN surrogate models are available at \url{https://github.com/chiyt14/SOEC-Homogeneity}.

\section*{Acknowledgements}
This work was supported by the National Natural Science Foundation of China [grant number 52177092];
Japan [grant number ]; 
The China Scholarship Council [grant number 202106210179].

\appendix

\setcounter{figure}{0}
\setcounter{table}{0}
\section{Graphical explanation of why a higher temperature results in a more inhomogeneous current distribution}
\label{sec:appendix1}
This section aims to provide a graphical explanation on the phenomenon demonstrated in Figure \ref{fig:Single-result}(c), Section \ref{sec:4-2}.
It says that the inhomogeneity of current distribution is enhanced when the temperature becomes higher along the contour line of $Q_{\rm H_2}=50$ sccm.
In other words, a higher $T_{\rm fur}$ leads to a higher $IH_{I}$.

\begin{figure}
  \centering
  \includegraphics[width=0.6\textwidth]{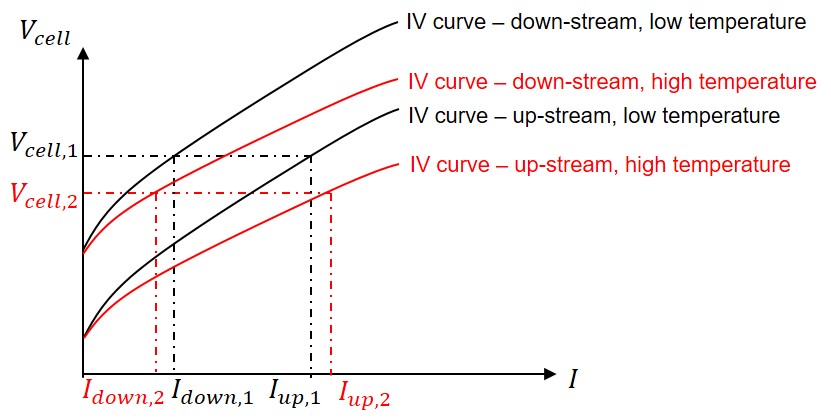}
  \caption{Qualitative explanation of why a higher $T_{\rm fur}$ leads to a higher $IH_{I}$.}
  \label{fig:GraphicalExp}
\end{figure}

A qualitative graphical explanation is given in Figure \ref{fig:GraphicalExp}.
Since $SU$ is fixed in Figure \ref{fig:Single-result}(c), Figure \ref{fig:GraphicalExp} demonstrates the IV curves of an SOEC with two segments when $SU$ is fixed.
No mass transfer limit can be observed considered because $SU$ is fixed.
Since the concentration distribution along the flow direction is almost determined when $SU$ is fixed, the difference between the up-stream IV curve and the down-stream IV curve is almost constant, owing to the Nernst loss caused by the decreasing steam concentration along the flow direction.
Therefore, the IV curves of the down-stream segment can be approximated by moving the IV curves of the up-stream segment in the upward direction.
When $T_{\rm fur}$ increases, the overpotential decreases, and the IV curves change from the black lines to the red lines.

Along the contour line of $Q_{\rm H_2}=50$ sccm in Figure \ref{fig:Single-result}(c), the total current is fixed.
In other words, $I_{\rm up,1}+I_{\rm down,1}=I_{\rm up,2}+I_{\rm down,2}$.
$I_{\rm up,1}$ and $I_{\rm up,2}$ (or $I_{\rm down,1}$ and $I_{\rm down,2}$) represent the currents of the up-stream (or down-stream) segment before and after the temperature rises.
Figure \ref{fig:GraphicalExp} shows that, to maintain a constant total current when the temperature rises, the cell voltage decreases from $V_{\rm cell,1}$ to $V_{\rm cell,2}$, while $I_{\rm up,1}<I_{\rm up,2}$ and $I_{\rm down,1}>I_{\rm down,2}$ hold.
Therefore, the inhomogeneity of the current distribution is enhanced, according to the definition of $IH_I$ in Equation (\ref{eqn:unindex}).
This phenomenon indicates that the current distribution becomes more inhomogeneous when the cell becomes more active.
This is only a qualitative explanation of the phenomenon.
A rigorous proof requires parametric expressions of the IV curves, which is not the focus of this paper.

\bibliographystyle{elsarticle-num-names}
\bibliography{references}

\begin{thebibliography}{41}
\expandafter\ifx\csname natexlab\endcsname\relax\def\natexlab#1{#1}\fi
\providecommand{\url}[1]{\texttt{#1}}
\providecommand{\href}[2]{#2}
\providecommand{\path}[1]{#1}
\providecommand{\DOIprefix}{doi:}
\providecommand{\ArXivprefix}{arXiv:}
\providecommand{\URLprefix}{URL: }
\providecommand{\Pubmedprefix}{pmid:}
\providecommand{\doi}[1]{\href{http://dx.doi.org/#1}{\path{#1}}}
\providecommand{\Pubmed}[1]{\href{pmid:#1}{\path{#1}}}
\providecommand{\bibinfo}[2]{#2}
\ifx\xfnm\relax \def\xfnm[#1]{\unskip,\space#1}\fi
\bibitem[{Wang et~al.(2020{\natexlab{a}})Wang, Deng, Zhang, Wang, Zhao, Miao,
  Yuan, and Yan}]{WANG2020-1}
\bibinfo{author}{F.~Wang}, \bibinfo{author}{S.~Deng},
  \bibinfo{author}{H.~Zhang}, \bibinfo{author}{J.~Wang},
  \bibinfo{author}{J.~Zhao}, \bibinfo{author}{H.~Miao},
  \bibinfo{author}{J.~Yuan}, \bibinfo{author}{J.~Yan},
\newblock \bibinfo{title}{A comprehensive review on high-temperature fuel cells
  with carbon capture},
\newblock \bibinfo{journal}{Applied Energy} \bibinfo{volume}{275}
  (\bibinfo{year}{2020}{\natexlab{a}}) \bibinfo{pages}{115342}. \URLprefix
  \url{https://www.sciencedirect.com/science/article/pii/S0306261920308540}.
  \DOIprefix\doi{https://doi.org/10.1016/j.apenergy.2020.115342}.
\bibitem[{Wang et~al.(2020{\natexlab{b}})Wang, Zhang, Pérez-Fortes, Aubin,
  Lin, Yang, Maréchal, and {Van herle}}]{WANG2020}
\bibinfo{author}{L.~Wang}, \bibinfo{author}{Y.~Zhang},
  \bibinfo{author}{M.~Pérez-Fortes}, \bibinfo{author}{P.~Aubin},
  \bibinfo{author}{T.-E. Lin}, \bibinfo{author}{Y.~Yang},
  \bibinfo{author}{F.~Maréchal}, \bibinfo{author}{J.~{Van herle}},
\newblock \bibinfo{title}{Reversible solid-oxide cell stack based
  power-to-x-to-power systems: Comparison of thermodynamic performance},
\newblock \bibinfo{journal}{Applied Energy} \bibinfo{volume}{275}
  (\bibinfo{year}{2020}{\natexlab{b}}) \bibinfo{pages}{115330}. \URLprefix
  \url{https://www.sciencedirect.com/science/article/pii/S0306261920308424}.
  \DOIprefix\doi{https://doi.org/10.1016/j.apenergy.2020.115330}.
\bibitem[{Frank et~al.(2018)Frank, Deja, Peters, Blum, and Stolten}]{FRANK2018}
\bibinfo{author}{M.~Frank}, \bibinfo{author}{R.~Deja},
  \bibinfo{author}{R.~Peters}, \bibinfo{author}{L.~Blum},
  \bibinfo{author}{D.~Stolten},
\newblock \bibinfo{title}{Bypassing renewable variability with a reversible
  solid oxide cell plant},
\newblock \bibinfo{journal}{Applied Energy} \bibinfo{volume}{217}
  (\bibinfo{year}{2018}) \bibinfo{pages}{101--112}. \URLprefix
  \url{https://www.sciencedirect.com/science/article/pii/S0306261918302460}.
  \DOIprefix\doi{https://doi.org/10.1016/j.apenergy.2018.02.115}.
\bibitem[{Khanafer et~al.(2022)Khanafer, Al-Masri, Vafai, and
  Preethichandra}]{KHANAFER2022}
\bibinfo{author}{K.~Khanafer}, \bibinfo{author}{A.~Al-Masri},
  \bibinfo{author}{K.~Vafai}, \bibinfo{author}{P.~Preethichandra},
\newblock \bibinfo{title}{Heat up impact on thermal stresses in sofc for mobile
  apu applications: Thermo-structural analysis},
\newblock \bibinfo{journal}{Sustainable Energy Technologies and Assessments}
  \bibinfo{volume}{52} (\bibinfo{year}{2022}) \bibinfo{pages}{102159}.
  \URLprefix
  \url{https://www.sciencedirect.com/science/article/pii/S2213138822002119}.
  \DOIprefix\doi{https://doi.org/10.1016/j.seta.2022.102159}.
\bibitem[{Xu et~al.(2021)Xu, Ma, Tan, Wu, Zhang, Ni, and Xuan}]{XU2021-2}
\bibinfo{author}{H.~Xu}, \bibinfo{author}{J.~Ma}, \bibinfo{author}{P.~Tan},
  \bibinfo{author}{Z.~Wu}, \bibinfo{author}{Y.~Zhang}, \bibinfo{author}{M.~Ni},
  \bibinfo{author}{J.~Xuan},
\newblock \bibinfo{title}{Enabling thermal-neutral electrolysis for co2-to-fuel
  conversions with a hybrid deep learning strategy},
\newblock \bibinfo{journal}{Energy Conversion and Management}
  \bibinfo{volume}{230} (\bibinfo{year}{2021}) \bibinfo{pages}{113827}.
  \URLprefix
  \url{https://www.sciencedirect.com/science/article/pii/S0196890421000042}.
  \DOIprefix\doi{https://doi.org/10.1016/j.enconman.2021.113827}.
\bibitem[{Damm and Fedorov(2006)}]{Damm2006}
\bibinfo{author}{D.~L. Damm}, \bibinfo{author}{A.~G. Fedorov},
\newblock \bibinfo{title}{Reduced-order transient thermal modeling for sofc
  heating and cooling},
\newblock \bibinfo{journal}{Journal of Power Sources} \bibinfo{volume}{159}
  (\bibinfo{year}{2006}) \bibinfo{pages}{956--967}. \URLprefix
  \url{https://www.sciencedirect.com/science/article/pii/S0378775305016393}.
  \DOIprefix\doi{https://doi.org/10.1016/j.jpowsour.2005.11.072}.
\bibitem[{Xinhai et~al.(2021)Xinhai, Yuhua, Zilin, and Zheng}]{Xu2021_1}
\bibinfo{author}{X.~Xinhai}, \bibinfo{author}{W.~Yuhua},
  \bibinfo{author}{Y.~Zilin}, \bibinfo{author}{Z.~Zheng},
\newblock \bibinfo{title}{Progress in macro scale mechanical effects
  investigation of solid oxide fuel cells},
\newblock \bibinfo{journal}{Advances in Mechanics} \bibinfo{volume}{51}
  (\bibinfo{year}{2021}) \bibinfo{pages}{62}.
  \DOIprefix\doi{10.6052/1000-0992-20-023}.
\bibitem[{Schiller et~al.(2009)Schiller, Bessler, Friedrich, Gewies, and
  Willich}]{Schiller_2009}
\bibinfo{author}{G.~Schiller}, \bibinfo{author}{W.~G. Bessler},
  \bibinfo{author}{K.~A. Friedrich}, \bibinfo{author}{S.~Gewies},
  \bibinfo{author}{C.~Willich},
\newblock \bibinfo{title}{Spatially resolved electrochemical performance in a
  segmented planar {SOFC}},
\newblock \bibinfo{journal}{{ECS} Transactions} \bibinfo{volume}{17}
  (\bibinfo{year}{2009}) \bibinfo{pages}{79--87}. \URLprefix
  \url{https://doi.org/10.1149/1.3142737}. \DOIprefix\doi{10.1149/1.3142737}.
\bibitem[{Canavar et~al.(2016)Canavar, Mat, Celik, Timurkutluk, and
  Kaplan}]{CANAVAR2016}
\bibinfo{author}{M.~Canavar}, \bibinfo{author}{A.~Mat},
  \bibinfo{author}{S.~Celik}, \bibinfo{author}{B.~Timurkutluk},
  \bibinfo{author}{Y.~Kaplan},
\newblock \bibinfo{title}{Investigation of temperature distribution and
  performance of sofc short stack with/without machined gas channels},
\newblock \bibinfo{journal}{International Journal of Hydrogen Energy}
  \bibinfo{volume}{41} (\bibinfo{year}{2016}) \bibinfo{pages}{10030--10036}.
  \URLprefix
  \url{https://www.sciencedirect.com/science/article/pii/S0360319916003190}.
  \DOIprefix\doi{https://doi.org/10.1016/j.ijhydene.2016.02.045},
  \bibinfo{note}{special Issue: 1st International Symposium on Materials for
  Energy Storage and Conversion (mESC-IS 2015)}.
\bibitem[{Razbani et~al.(2013)Razbani, Wærnhus, and Assadi}]{RAZBANI2013}
\bibinfo{author}{O.~Razbani}, \bibinfo{author}{I.~Wærnhus},
  \bibinfo{author}{M.~Assadi},
\newblock \bibinfo{title}{Experimental investigation of temperature
  distribution over a planar solid oxide fuel cell},
\newblock \bibinfo{journal}{Applied Energy} \bibinfo{volume}{105}
  (\bibinfo{year}{2013}) \bibinfo{pages}{155--160}. \URLprefix
  \url{https://www.sciencedirect.com/science/article/pii/S0306261912009531}.
  \DOIprefix\doi{https://doi.org/10.1016/j.apenergy.2012.12.062}.
\bibitem[{Wu et~al.(2022)Wu, Liu, Wang, An, and Xu}]{WU2022}
\bibinfo{author}{Y.~Wu}, \bibinfo{author}{H.~Liu}, \bibinfo{author}{Y.~Wang},
  \bibinfo{author}{L.~An}, \bibinfo{author}{X.~Xu},
\newblock \bibinfo{title}{Spatially resolved electrochemical performance and
  temperature distribution of a segmented solid oxide fuel cell under various
  hydrogen dilution ratios and electrical loadings},
\newblock \bibinfo{journal}{Journal of Power Sources} \bibinfo{volume}{536}
  (\bibinfo{year}{2022}) \bibinfo{pages}{231477}. \URLprefix
  \url{https://www.sciencedirect.com/science/article/pii/S0378775322004840}.
  \DOIprefix\doi{https://doi.org/10.1016/j.jpowsour.2022.231477}.
\bibitem[{Guk et~al.(2019)Guk, Venkatesan, Babar, Jackson, and Kim}]{GUK2019}
\bibinfo{author}{E.~Guk}, \bibinfo{author}{V.~Venkatesan},
  \bibinfo{author}{S.~Babar}, \bibinfo{author}{L.~Jackson},
  \bibinfo{author}{J.-S. Kim},
\newblock \bibinfo{title}{Parameters and their impacts on the temperature
  distribution and thermal gradient of solid oxide fuel cell},
\newblock \bibinfo{journal}{Applied Energy} \bibinfo{volume}{241}
  (\bibinfo{year}{2019}) \bibinfo{pages}{164--173}. \URLprefix
  \url{https://www.sciencedirect.com/science/article/pii/S0306261919304337}.
  \DOIprefix\doi{https://doi.org/10.1016/j.apenergy.2019.03.034}.
\bibitem[{Sugihara and Iwai(2021)}]{SUGIHARA2021}
\bibinfo{author}{S.~Sugihara}, \bibinfo{author}{H.~Iwai},
\newblock \bibinfo{title}{Measurement of transient temperature distribution
  behavior of a planar solid oxide fuel cell: Effect of instantaneous switching
  of power generation and direct internal reforming},
\newblock \bibinfo{journal}{Journal of Power Sources} \bibinfo{volume}{482}
  (\bibinfo{year}{2021}) \bibinfo{pages}{229070}. \URLprefix
  \url{https://www.sciencedirect.com/science/article/pii/S0378775320313653}.
  \DOIprefix\doi{https://doi.org/10.1016/j.jpowsour.2020.229070}.
\bibitem[{Sugihara and Iwai(2020)}]{SUGIHARA2020}
\bibinfo{author}{S.~Sugihara}, \bibinfo{author}{H.~Iwai},
\newblock \bibinfo{title}{Experimental investigation of temperature
  distribution of planar solid oxide fuel cell: Effects of gas flow, power
  generation, and direct internal reforming},
\newblock \bibinfo{journal}{International Journal of Hydrogen Energy}
  \bibinfo{volume}{45} (\bibinfo{year}{2020}) \bibinfo{pages}{25227--25239}.
  \URLprefix
  \url{https://www.sciencedirect.com/science/article/pii/S0360319920321704}.
  \DOIprefix\doi{https://doi.org/10.1016/j.ijhydene.2020.06.033}.
\bibitem[{Eigenbrodt et~al.(2011)Eigenbrodt, Pomfret, Steinhurst, Owrutsky, and
  Walker}]{Eigenbrodt2011}
\bibinfo{author}{B.~C. Eigenbrodt}, \bibinfo{author}{M.~B. Pomfret},
  \bibinfo{author}{D.~A. Steinhurst}, \bibinfo{author}{J.~C. Owrutsky},
  \bibinfo{author}{R.~A. Walker},
\newblock \bibinfo{title}{Direct, in situ optical studies of ni-ysz anodes in
  solid oxide fuel cells operating with methanol and methane},
\newblock \bibinfo{journal}{The Journal of Physical Chemistry C}
  \bibinfo{volume}{115} (\bibinfo{year}{2011}) \bibinfo{pages}{2895--2903}.
  \DOIprefix\doi{10.1021/jp109292r}.
\bibitem[{Zaghloul et~al.(2021)Zaghloul, Mason, Wang, Buric, Peng, Lee,
  Ohodnicki, Abernathy, and Chen}]{ZAGHLOUL2021}
\bibinfo{author}{M.~A. Zaghloul}, \bibinfo{author}{J.~H. Mason},
  \bibinfo{author}{M.~Wang}, \bibinfo{author}{M.~Buric},
  \bibinfo{author}{Z.~Peng}, \bibinfo{author}{S.~Lee},
  \bibinfo{author}{P.~Ohodnicki}, \bibinfo{author}{H.~Abernathy},
  \bibinfo{author}{K.~P. Chen},
\newblock \bibinfo{title}{High spatial resolution temperature profile
  measurements of solid-oxide fuel cells},
\newblock \bibinfo{journal}{Applied Energy} \bibinfo{volume}{288}
  (\bibinfo{year}{2021}) \bibinfo{pages}{116633}. \URLprefix
  \url{https://www.sciencedirect.com/science/article/pii/S0306261921001689}.
  \DOIprefix\doi{https://doi.org/10.1016/j.apenergy.2021.116633}.
\bibitem[{Zhang et~al.(2010)Zhang, Guo, Ma, and Liu}]{ZHANG2010}
\bibinfo{author}{G.~Zhang}, \bibinfo{author}{L.~Guo}, \bibinfo{author}{L.~Ma},
  \bibinfo{author}{H.~Liu},
\newblock \bibinfo{title}{Simultaneous measurement of current and temperature
  distributions in a proton exchange membrane fuel cell},
\newblock \bibinfo{journal}{Journal of Power Sources} \bibinfo{volume}{195}
  (\bibinfo{year}{2010}) \bibinfo{pages}{3597--3604}. \URLprefix
  \url{https://www.sciencedirect.com/science/article/pii/S037877530902223X}.
  \DOIprefix\doi{https://doi.org/10.1016/j.jpowsour.2009.12.016}.
\bibitem[{Wang et~al.(2020)Wang, Xie, Zhou, Feng, Zhou, Yuan, Xu, Fan, Zeng,
  Li, and Wang}]{WANG2020-2}
\bibinfo{author}{Y.~Wang}, \bibinfo{author}{X.~Xie}, \bibinfo{author}{C.~Zhou},
  \bibinfo{author}{Q.~Feng}, \bibinfo{author}{Y.~Zhou}, \bibinfo{author}{X.-Z.
  Yuan}, \bibinfo{author}{J.~Xu}, \bibinfo{author}{J.~Fan},
  \bibinfo{author}{L.~Zeng}, \bibinfo{author}{H.~Li},
  \bibinfo{author}{H.~Wang},
\newblock \bibinfo{title}{Study of relative humidity on durability of the
  reversal tolerant proton exchange membrane fuel cell anode using a segmented
  cell},
\newblock \bibinfo{journal}{Journal of Power Sources} \bibinfo{volume}{449}
  (\bibinfo{year}{2020}) \bibinfo{pages}{227542}. \URLprefix
  \url{https://www.sciencedirect.com/science/article/pii/S0378775319315356}.
  \DOIprefix\doi{https://doi.org/10.1016/j.jpowsour.2019.227542}.
\bibitem[{Wuillemin et~al.(2008)Wuillemin, M{\"u}ller, Nakajo, Autissier,
  Diethelm, Molinelli, Favrat et~al.}]{wuillemin2008}
\bibinfo{author}{Z.~Wuillemin}, \bibinfo{author}{A.~M{\"u}ller},
  \bibinfo{author}{A.~Nakajo}, \bibinfo{author}{N.~Autissier},
  \bibinfo{author}{S.~Diethelm}, \bibinfo{author}{M.~Molinelli},
  \bibinfo{author}{D.~Favrat}, et~al.,
\newblock \bibinfo{title}{Investigation of local electrochemical performance
  and local degradation in an operating solid oxide fuel cell},
\newblock in: \bibinfo{booktitle}{Proceedings of the 8th EUROPEAN FUEL CELL
  FORUM 2008}, \bibinfo{number}{CONF}, \bibinfo{organization}{European Fuel
  Cell Forum}, \bibinfo{year}{2008}, pp. \bibinfo{pages}{1--20}.
\bibitem[{Özgür Aydın et~al.(2015{\natexlab{a}})Özgür Aydın, Koshiyama,
  Nakajima, and Kitahara}]{AYDIN2015}
\bibinfo{author}{Özgür Aydın}, \bibinfo{author}{T.~Koshiyama},
  \bibinfo{author}{H.~Nakajima}, \bibinfo{author}{T.~Kitahara},
\newblock \bibinfo{title}{In-situ diagnosis and assessment of longitudinal
  current variation by electrode-segmentation method in anode-supported
  microtubular solid oxide fuel cells},
\newblock \bibinfo{journal}{Journal of Power Sources} \bibinfo{volume}{279}
  (\bibinfo{year}{2015}{\natexlab{a}}) \bibinfo{pages}{218--223}. \URLprefix
  \url{https://www.sciencedirect.com/science/article/pii/S0378775314021958}.
  \DOIprefix\doi{https://doi.org/10.1016/j.jpowsour.2014.12.156},
  \bibinfo{note}{9th International Conference on Lead-Acid Batteries – LABAT
  2014}.
\bibitem[{Özgür Aydın et~al.(2015{\natexlab{b}})Özgür Aydın, Nakajima,
  and Kitahara}]{AYDIN2015-1}
\bibinfo{author}{Özgür Aydın}, \bibinfo{author}{H.~Nakajima},
  \bibinfo{author}{T.~Kitahara},
\newblock \bibinfo{title}{Current and temperature distributions in-situ
  acquired by electrode-segmentation along a microtubular solid oxide fuel cell
  operating with syngas},
\newblock \bibinfo{journal}{Journal of Power Sources} \bibinfo{volume}{293}
  (\bibinfo{year}{2015}{\natexlab{b}}) \bibinfo{pages}{1053--1061}. \URLprefix
  \url{https://www.sciencedirect.com/science/article/pii/S0378775315010587}.
  \DOIprefix\doi{https://doi.org/10.1016/j.jpowsour.2015.06.024}.
\bibitem[{Kim et~al.(2020)Kim, Son, Park, Park, and Lim}]{KIM2019}
\bibinfo{author}{M.~W. Kim}, \bibinfo{author}{M.~J. Son},
  \bibinfo{author}{H.~Park}, \bibinfo{author}{J.-Y. Park},
  \bibinfo{author}{H.-T. Lim},
\newblock \bibinfo{title}{Experimental investigation of in-plane performance
  variation on anode supported solid oxide fuel cells using segmented cathodes
  and reference electrodes},
\newblock \bibinfo{journal}{Fuel Cells} \bibinfo{volume}{20}
  (\bibinfo{year}{2020}) \bibinfo{pages}{212--219}. \URLprefix
  \url{https://onlinelibrary.wiley.com/doi/abs/10.1002/fuce.201900214}.
  \DOIprefix\doi{https://doi.org/10.1002/fuce.201900214}.
  \href{http://arxiv.org/abs/https://onlinelibrary.wiley.com/doi/pdf/10.1002/fuce.201900214}{{\tt
  arXiv:https://onlinelibrary.wiley.com/doi/pdf/10.1002/fuce.201900214}}.
\bibitem[{Özgür Aydın et~al.(2016)Özgür Aydın, Nakajima, and
  Kitahara}]{AYDIN2016}
\bibinfo{author}{Özgür Aydın}, \bibinfo{author}{H.~Nakajima},
  \bibinfo{author}{T.~Kitahara},
\newblock \bibinfo{title}{Reliability of the numerical sofc models for
  estimating the spatial current and temperature variations},
\newblock \bibinfo{journal}{International Journal of Hydrogen Energy}
  \bibinfo{volume}{41} (\bibinfo{year}{2016}) \bibinfo{pages}{15311--15324}.
  \URLprefix
  \url{https://www.sciencedirect.com/science/article/pii/S0360319916319000}.
  \DOIprefix\doi{https://doi.org/10.1016/j.ijhydene.2016.06.194}.
\bibitem[{Bessler et~al.(2010)Bessler, Gewies, Willich, Schiller, and
  Friedrich}]{Bessler2010}
\bibinfo{author}{W.~G. Bessler}, \bibinfo{author}{S.~Gewies},
  \bibinfo{author}{C.~Willich}, \bibinfo{author}{G.~Schiller},
  \bibinfo{author}{K.~A. Friedrich},
\newblock \bibinfo{title}{Spatial distribution of electrochemical performance
  in a segmented sofc: A combined modeling and experimental study},
\newblock \bibinfo{journal}{Fuel Cells} \bibinfo{volume}{10}
  (\bibinfo{year}{2010}) \bibinfo{pages}{411--418}. \URLprefix
  \url{https://onlinelibrary.wiley.com/doi/abs/10.1002/fuce.200900083}.
  \DOIprefix\doi{https://doi.org/10.1002/fuce.200900083}.
  \href{http://arxiv.org/abs/https://onlinelibrary.wiley.com/doi/pdf/10.1002/fuce.200900083}{{\tt
  arXiv:https://onlinelibrary.wiley.com/doi/pdf/10.1002/fuce.200900083}}.
\bibitem[{Özgür Aydın et~al.(2018)Özgür Aydın, Ochiai, Nakajima,
  Kitahara, Ito, Ogura, and Shimano}]{AYDIN2018}
\bibinfo{author}{Özgür Aydın}, \bibinfo{author}{T.~Ochiai},
  \bibinfo{author}{H.~Nakajima}, \bibinfo{author}{T.~Kitahara},
  \bibinfo{author}{K.~Ito}, \bibinfo{author}{Y.~Ogura},
  \bibinfo{author}{J.~Shimano},
\newblock \bibinfo{title}{Mass transport limitation in inlet periphery of fuel
  cells: Studied on a planar solid oxide fuel cell},
\newblock \bibinfo{journal}{International Journal of Hydrogen Energy}
  \bibinfo{volume}{43} (\bibinfo{year}{2018}) \bibinfo{pages}{17420--17430}.
  \URLprefix
  \url{https://www.sciencedirect.com/science/article/pii/S0360319918321517}.
  \DOIprefix\doi{https://doi.org/10.1016/j.ijhydene.2018.07.030}.
\bibitem[{Arriagada et~al.(2002)Arriagada, Olausson, and
  Selimovic}]{ARRIAGADA2002}
\bibinfo{author}{J.~Arriagada}, \bibinfo{author}{P.~Olausson},
  \bibinfo{author}{A.~Selimovic},
\newblock \bibinfo{title}{Artificial neural network simulator for sofc
  performance prediction},
\newblock \bibinfo{journal}{Journal of Power Sources} \bibinfo{volume}{112}
  (\bibinfo{year}{2002}) \bibinfo{pages}{54--60}. \URLprefix
  \url{https://www.sciencedirect.com/science/article/pii/S0378775302003142}.
  \DOIprefix\doi{https://doi.org/10.1016/S0378-7753(02)00314-2}.
\bibitem[{Huo et~al.(2006)Huo, Zhu, and Cao}]{HUO2006}
\bibinfo{author}{H.-B. Huo}, \bibinfo{author}{X.-J. Zhu},
  \bibinfo{author}{G.-Y. Cao},
\newblock \bibinfo{title}{Nonlinear modeling of a sofc stack based on a least
  squares support vector machine},
\newblock \bibinfo{journal}{Journal of Power Sources} \bibinfo{volume}{162}
  (\bibinfo{year}{2006}) \bibinfo{pages}{1220--1225}. \URLprefix
  \url{https://www.sciencedirect.com/science/article/pii/S0378775306012870}.
  \DOIprefix\doi{https://doi.org/10.1016/j.jpowsour.2006.07.031},
  \bibinfo{note}{special issue including selected papers from the International
  Power Sources Symposium 2005 together with regular papers}.
\bibitem[{Zahadat and Milewski(2015)}]{ZAHADAT2015}
\bibinfo{author}{P.~Zahadat}, \bibinfo{author}{J.~Milewski},
\newblock \bibinfo{title}{Modeling electrical behavior of solid oxide
  electrolyzer cells by using artificial neural network},
\newblock \bibinfo{journal}{International Journal of Hydrogen Energy}
  \bibinfo{volume}{40} (\bibinfo{year}{2015}) \bibinfo{pages}{7246--7251}.
  \URLprefix
  \url{https://www.sciencedirect.com/science/article/pii/S0360319915009301}.
  \DOIprefix\doi{https://doi.org/10.1016/j.ijhydene.2015.04.042}.
\bibitem[{Milewski and Świrski(2009)}]{MILEWSKI2009}
\bibinfo{author}{J.~Milewski}, \bibinfo{author}{K.~Świrski},
\newblock \bibinfo{title}{Modelling the sofc behaviours by artificial neural
  network},
\newblock \bibinfo{journal}{International Journal of Hydrogen Energy}
  \bibinfo{volume}{34} (\bibinfo{year}{2009}) \bibinfo{pages}{5546--5553}.
  \URLprefix
  \url{https://www.sciencedirect.com/science/article/pii/S0360319909006405}.
  \DOIprefix\doi{https://doi.org/10.1016/j.ijhydene.2009.04.068}.
\bibitem[{Chi et~al.(2020)Chi, Qiu, Lin, Song, Li, Hu, Mu, and Liu}]{CHI2020}
\bibinfo{author}{Y.~Chi}, \bibinfo{author}{Y.~Qiu}, \bibinfo{author}{J.~Lin},
  \bibinfo{author}{Y.~Song}, \bibinfo{author}{W.~Li}, \bibinfo{author}{Q.~Hu},
  \bibinfo{author}{S.~Mu}, \bibinfo{author}{M.~Liu},
\newblock \bibinfo{title}{A robust surrogate model of a solid oxide cell based
  on an adaptive polynomial approximation method},
\newblock \bibinfo{journal}{International Journal of Hydrogen Energy}
  \bibinfo{volume}{45} (\bibinfo{year}{2020}) \bibinfo{pages}{32949--32971}.
  \URLprefix
  \url{https://www.sciencedirect.com/science/article/pii/S0360319920335564}.
  \DOIprefix\doi{https://doi.org/10.1016/j.ijhydene.2020.09.116}.
\bibitem[{Cai et~al.(2014)Cai, Adjiman, and Brandon}]{CAI2014}
\bibinfo{author}{Q.~Cai}, \bibinfo{author}{C.~S. Adjiman},
  \bibinfo{author}{N.~P. Brandon},
\newblock \bibinfo{title}{Optimal control strategies for hydrogen production
  when coupling solid oxide electrolysers with intermittent renewable
  energies},
\newblock \bibinfo{journal}{Journal of Power Sources} \bibinfo{volume}{268}
  (\bibinfo{year}{2014}) \bibinfo{pages}{212--224}. \URLprefix
  \url{https://www.sciencedirect.com/science/article/pii/S0378775314008817}.
  \DOIprefix\doi{https://doi.org/10.1016/j.jpowsour.2014.06.028}.
\bibitem[{Xing et~al.(2018)Xing, Lin, Song, Hu, Zhou, and Mu}]{XING2018}
\bibinfo{author}{X.~Xing}, \bibinfo{author}{J.~Lin}, \bibinfo{author}{Y.~Song},
  \bibinfo{author}{Q.~Hu}, \bibinfo{author}{Y.~Zhou}, \bibinfo{author}{S.~Mu},
\newblock \bibinfo{title}{Optimization of hydrogen yield of a high-temperature
  electrolysis system with coordinated temperature and feed factors at various
  loading conditions: A model-based study},
\newblock \bibinfo{journal}{Applied Energy} \bibinfo{volume}{232}
  (\bibinfo{year}{2018}) \bibinfo{pages}{368--385}. \URLprefix
  \url{https://www.sciencedirect.com/science/article/pii/S030626191831328X}.
  \DOIprefix\doi{https://doi.org/10.1016/j.apenergy.2018.09.020}.
\bibitem[{Xu et~al.(2020)Xu, Ma, Tan, Chen, Wu, Zhang, Wang, Xuan, and
  Ni}]{XU2020}
\bibinfo{author}{H.~Xu}, \bibinfo{author}{J.~Ma}, \bibinfo{author}{P.~Tan},
  \bibinfo{author}{B.~Chen}, \bibinfo{author}{Z.~Wu},
  \bibinfo{author}{Y.~Zhang}, \bibinfo{author}{H.~Wang},
  \bibinfo{author}{J.~Xuan}, \bibinfo{author}{M.~Ni},
\newblock \bibinfo{title}{Towards online optimisation of solid oxide fuel cell
  performance: Combining deep learning with multi-physics simulation},
\newblock \bibinfo{journal}{Energy and AI} \bibinfo{volume}{1}
  (\bibinfo{year}{2020}) \bibinfo{pages}{100003}. \URLprefix
  \url{https://www.sciencedirect.com/science/article/pii/S2666546820300033}.
  \DOIprefix\doi{https://doi.org/10.1016/j.egyai.2020.100003}.
\bibitem[{Sun et~al.(2022)Sun, Lu, Liu, Shuai, Sun, Zheng, Han, Xiao, Xuan, Ni,
  and Xu}]{SUN2022}
\bibinfo{author}{Y.~Sun}, \bibinfo{author}{J.~Lu}, \bibinfo{author}{Q.~Liu},
  \bibinfo{author}{W.~Shuai}, \bibinfo{author}{A.~Sun},
  \bibinfo{author}{N.~Zheng}, \bibinfo{author}{Y.~Han},
  \bibinfo{author}{G.~Xiao}, \bibinfo{author}{J.~Xuan},
  \bibinfo{author}{M.~Ni}, \bibinfo{author}{H.~Xu},
\newblock \bibinfo{title}{Multi-objective optimizations of solid oxide
  co-electrolysis with intermittent renewable power supply via multi-physics
  simulation and deep learning strategy},
\newblock \bibinfo{journal}{Energy Conversion and Management}
  \bibinfo{volume}{258} (\bibinfo{year}{2022}) \bibinfo{pages}{115560}.
  \URLprefix
  \url{https://www.sciencedirect.com/science/article/pii/S0196890422003569}.
  \DOIprefix\doi{https://doi.org/10.1016/j.enconman.2022.115560}.
\bibitem[{Wehrle et~al.(2022)Wehrle, Schmider, Dailly, Banerjee, and
  Deutschmann}]{WEHRLE2022}
\bibinfo{author}{L.~Wehrle}, \bibinfo{author}{D.~Schmider},
  \bibinfo{author}{J.~Dailly}, \bibinfo{author}{A.~Banerjee},
  \bibinfo{author}{O.~Deutschmann},
\newblock \bibinfo{title}{Benchmarking solid oxide electrolysis cell-stacks for
  industrial power-to-methane systems via hierarchical multi-scale modelling},
\newblock \bibinfo{journal}{Applied Energy} \bibinfo{volume}{317}
  (\bibinfo{year}{2022}) \bibinfo{pages}{119143}. \URLprefix
  \url{https://www.sciencedirect.com/science/article/pii/S0306261922005190}.
  \DOIprefix\doi{https://doi.org/10.1016/j.apenergy.2022.119143}.
\bibitem[{Yan et~al.(2019)Yan, He, Hara, and Shikazono}]{YAN2019}
\bibinfo{author}{Z.~Yan}, \bibinfo{author}{A.~He}, \bibinfo{author}{S.~Hara},
  \bibinfo{author}{N.~Shikazono},
\newblock \bibinfo{title}{Modeling of solid oxide fuel cell (sofc) electrodes
  from fabrication to operation: Microstructure optimization via artificial
  neural networks and multi-objective genetic algorithms},
\newblock \bibinfo{journal}{Energy Conversion and Management}
  \bibinfo{volume}{198} (\bibinfo{year}{2019}) \bibinfo{pages}{111916}.
  \URLprefix
  \url{https://www.sciencedirect.com/science/article/pii/S0196890419309070}.
  \DOIprefix\doi{https://doi.org/10.1016/j.enconman.2019.111916}.
\bibitem[{Saltelli et~al.(2010)Saltelli, Annoni, Azzini, Campolongo, Ratto, and
  Tarantola}]{SALTELLI2010}
\bibinfo{author}{A.~Saltelli}, \bibinfo{author}{P.~Annoni},
  \bibinfo{author}{I.~Azzini}, \bibinfo{author}{F.~Campolongo},
  \bibinfo{author}{M.~Ratto}, \bibinfo{author}{S.~Tarantola},
\newblock \bibinfo{title}{Variance based sensitivity analysis of model output.
  design and estimator for the total sensitivity index},
\newblock \bibinfo{journal}{Computer Physics Communications}
  \bibinfo{volume}{181} (\bibinfo{year}{2010}) \bibinfo{pages}{259--270}.
  \URLprefix
  \url{https://www.sciencedirect.com/science/article/pii/S0010465509003087}.
  \DOIprefix\doi{https://doi.org/10.1016/j.cpc.2009.09.018}.
\bibitem[{Sharifzadeh et~al.(2017)Sharifzadeh, Meghdari, and
  Rashtchian}]{SHARIFZADEH2017}
\bibinfo{author}{M.~Sharifzadeh}, \bibinfo{author}{M.~Meghdari},
  \bibinfo{author}{D.~Rashtchian},
\newblock \bibinfo{title}{Multi-objective design and operation of solid oxide
  fuel cell (sofc) triple combined-cycle power generation systems: Integrating
  energy efficiency and operational safety},
\newblock \bibinfo{journal}{Applied Energy} \bibinfo{volume}{185}
  (\bibinfo{year}{2017}) \bibinfo{pages}{345--361}. \URLprefix
  \url{https://www.sciencedirect.com/science/article/pii/S0306261916315963}.
  \DOIprefix\doi{https://doi.org/10.1016/j.apenergy.2016.11.010}.
\bibitem[{Behzadi et~al.(2021)Behzadi, Habibollahzade, Arabkoohsar, Shabani,
  Fakhari, and Vojdani}]{BEHZADI2021}
\bibinfo{author}{A.~Behzadi}, \bibinfo{author}{A.~Habibollahzade},
  \bibinfo{author}{A.~Arabkoohsar}, \bibinfo{author}{B.~Shabani},
  \bibinfo{author}{I.~Fakhari}, \bibinfo{author}{M.~Vojdani},
\newblock \bibinfo{title}{4e analysis of efficient waste heat recovery from
  sofc using apc: An effort to reach maximum efficiency and minimum emission
  through an application of grey wolf optimization},
\newblock \bibinfo{journal}{International Journal of Hydrogen Energy}
  \bibinfo{volume}{46} (\bibinfo{year}{2021}) \bibinfo{pages}{23879--23897}.
  \URLprefix
  \url{https://www.sciencedirect.com/science/article/pii/S0360319921016347}.
  \DOIprefix\doi{https://doi.org/10.1016/j.ijhydene.2021.04.187}.
\bibitem[{Lei et~al.(2022)Lei, Ye, Xu, Kong, Xu, Chen, Huang, and
  Xiao}]{LEI2022}
\bibinfo{author}{Y.~Lei}, \bibinfo{author}{S.~Ye}, \bibinfo{author}{Y.~Xu},
  \bibinfo{author}{C.~Kong}, \bibinfo{author}{C.~Xu},
  \bibinfo{author}{Y.~Chen}, \bibinfo{author}{W.~Huang},
  \bibinfo{author}{H.~Xiao},
\newblock \bibinfo{title}{Multi-objective optimization and algorithm
  improvement on thermal coupling of sofc-gt-orc integrated system},
\newblock \bibinfo{journal}{Computers \& Chemical Engineering}
  \bibinfo{volume}{164} (\bibinfo{year}{2022}) \bibinfo{pages}{107903}.
  \URLprefix
  \url{https://www.sciencedirect.com/science/article/pii/S0098135422002411}.
  \DOIprefix\doi{https://doi.org/10.1016/j.compchemeng.2022.107903}.
\bibitem[{Zhu et~al.(2021)Zhu, Wu, Guo, Yao, Dai, Ren, Kurko, Yan, Yang, and
  Zhang}]{ZHU2021}
\bibinfo{author}{P.~Zhu}, \bibinfo{author}{Z.~Wu}, \bibinfo{author}{L.~Guo},
  \bibinfo{author}{J.~Yao}, \bibinfo{author}{M.~Dai}, \bibinfo{author}{J.~Ren},
  \bibinfo{author}{S.~Kurko}, \bibinfo{author}{H.~Yan},
  \bibinfo{author}{F.~Yang}, \bibinfo{author}{Z.~Zhang},
\newblock \bibinfo{title}{Achieving high-efficiency conversion and
  poly-generation of cooling, heating, and power based on biomass-fueled sofc
  hybrid system: Performance assessment and multi-objective optimization},
\newblock \bibinfo{journal}{Energy Conversion and Management}
  \bibinfo{volume}{240} (\bibinfo{year}{2021}) \bibinfo{pages}{114245}.
  \URLprefix
  \url{https://www.sciencedirect.com/science/article/pii/S0196890421004210}.
  \DOIprefix\doi{https://doi.org/10.1016/j.enconman.2021.114245}.

\end{thebibliography}

\end{document}